\documentclass[11pt,a4paper]{article}
\usepackage{amsfonts,amsmath,amssymb}
\usepackage[title,titletoc]{appendix}
\usepackage{graphics,epsfig,epstopdf}

\newcommand{\foot}{\footnote}
\newcommand{\bi}{\bibitem}
\newcommand{\ci}{\cite}
\newcommand{\rf}{\eqref}

\newcommand{\fr}{\frac}

\newcommand{\x}{\times}

\newcommand{\cosech}{\textrm{\,cosech\,}}

\newcommand{\del}{\partial}
\newcommand{\dpl}{\partial_+}
\newcommand{\dm}{\partial_-}

\newcommand{\dpm}{\partial_\pm}

\newcommand{\ra}{\rightarrow}

\newcommand{\id}{\tbf{1}}
\newcommand{\inv}[1]{#1^{-1}}
\newcommand{\ha}{\fr{1}{2}}
\newcommand{\nb}[1]{{(#1)}}

\newcommand{\td}{\tilde}

\newcommand{\al}{\alpha}
\newcommand{\bet}{\beta}
\newcommand{\g}{\gamma}
\newcommand{\de}{\delta}
\newcommand{\De}{\Delta}

\newcommand{\ep}{\epsilon}

\newcommand{\ze}{\zeta}
\newcommand{\h}{\eta}
\newcommand{\q}{\theta}

\newcommand{\s}{\sigma}

\newcommand{\ta}{\tau}

\newcommand{\be}{\begin{equation}}
\newcommand{\ee}{\end{equation}}
\newcommand{\bd}{\begin{displaymath}}
\newcommand{\ed}{\end{displaymath}}
\newcommand{\ba}{\begin{array}}
\newcommand{\ea}{\end{array}}
\newcommand{\bea}{\begin{eqnarray}}
\newcommand{\eea}{\end{eqnarray}}
\newcommand{\bs}{\begin{split}}
\newcommand{\bp}{\begin{pmatrix}} 
\newcommand{\emp}{\end{pmatrix}}
\newcommand{\no}{\nonumber}
\newcommand{\la}{\label}
\newcommand{\bit}{\begin{itemize}}
\newcommand{\eit}{\end{itemize}}

\newcommand{\hs}{\hspace}
\newcommand{\vs}{\vspace}

\newcommand{\ts}{\thinspace}

\newcommand{\mc}{\mathcal}
\newcommand{\mf}{\mathfrak}
\newcommand{\mbb}{\mathbb}
\newcommand{\trm}{\textrm}
\newcommand{\tbf}{\textbf}

\newcommand{\Tr}{\trm{Tr}}
\newcommand{\tr}{\trm{tr}}

\newcommand{\com}[2]{[ #1,\,#2 ]}

\newcommand{\ord}[1]{\mathcal{O}(#1)}

\newcommand{\Lag}{\mathcal{L}}

\newcommand{\bra}[1]{\left<#1\right|}
\newcommand{\ket}[1]{\left|#1\right>}

\newcommand{\ald}{\dot{\al}}
\newcommand{\betd}{\dot{\bet}}
\newcommand{\gad}{\dot{\g}}
\newcommand{\ded}{\dot{\de}}

\newcommand{\su}{\mf{su}}
\newcommand{\so}{\mf{so}}

\newcommand{\adss}{$AdS_5 \x S^5$ }

\newcommand{\ind}{\nu}
\newcommand{\dcn}{{c_{_G}}}
\newcommand{\TT}{{\rm T}}
\newcommand{\TTm}{{\rm T}}
\newcommand{\hS}{\hat{S}} 
\newcommand{\mQ}{\mc{Q}}

\newcommand{\zg}{{\rm g}}
\newcommand{\gok}{m^2}
\def \sign {{\rm sign}}
\def \M {{\rm M}}

\begin{document}
\thispagestyle{empty}

\vs{-1cm}
\rightline{Imperial-TP-BH-2010-01}

\begin{center}
\vskip 1cm
{\Large\bf On  the perturbative S-matrix of 
        generalized sine-Gordon models}
\vskip 1.2cm 
{B. Hoare\foot{benjamin.hoare08@imperial.ac.uk}
 and A.A. Tseytlin\foot{Also at Lebedev Institute, Moscow.
 tseytlin@imperial.ac.uk}}
\\
\vskip 0.6cm
{\em Theoretical Physics Group \\
Blackett Laboratory, Imperial College \\
London, SW7 2AZ, U.K.}
\end{center}

\begin{abstract} \noindent 
Motivated by its relation to the Pohlmeyer reduction 
of $AdS_5 \times S^5$ superstring 
theory we continue
 the investigation of the 
 generalized sine-Gordon model 
defined by $SO(N+1)/SO(N)$ gauged WZW theory  with an 
 integrable potential. Extending our previous work (arXiv:0912.2958) 
we compute the  one-loop two-particle S-matrix for the elementary massive
excitations. In the $N=2$ case
corresponding to the  complex sine-Gordon theory  it agrees with the 
charge-one sector of the  quantum soliton S-matrix proposed in hep-th/9410140.
In the case of $N > 2$  when the gauge group $SO(N)$ is non-abelian 
we find a curious  anomaly in the Yang-Baxter equation which we 
interpret as a gauge artifact related to the fact 
that the scattered  particles are not singlets 
under the residual global subgroup of the gauge group. 
\end{abstract}

\setcounter{footnote}{0}
\newpage

\vspace{2cm}
\tableofcontents
\newpage

\renewcommand{\theequation}{1.\arabic{equation}}
\setcounter{equation}{0}

\section{Introduction\la{intro}}

In this paper we continue the investigation \ci{ht} of the 
perturbative S-matrix of generalised sine-Gordon models. 
Various examples of such   models (called also ``symmetric space
sine-Gordon models'')
based  on a $G/H$ gauged WZW   model with an integrable
 potential term 
were considered in, e.g.,  \ci{pol,bakas,mist,bakps,fpgm}.

The action for these $G/H$ models is  given by the (asymmetrically)
gauged WZW  action with a  potential  term, 
\be\la{gwzwact}\begin{split}
\mc{S} & 
 = - \ \fr{k}{8\pi\ind}  \trm{Tr} \Big[ \fr{1}{2}\,\int d^2x
\; 
      \ \inv{g}\dpl g\ \inv{g}\dm g\  
    - \fr{1}{3}\,\int d^3x \;
      \ \epsilon^{mnl}  
      \ \inv{g} \del_m g\ \inv{g}\del_n g \ \inv{g}\del_l g
\\     & \qquad\quad +  \,\int d^2x \;
      \ \big( A_+\dm g\inv{g} - A_-\inv{g}\dpl g 
             - \inv{g} A_+g A_-  +  \tau(A_+)A_- \big)
\\     & \qquad\quad + \gok\,\int d^2x\; 
      \ (\inv{g} T g T - T^2) \Big]\,.
\end{split}\ee
Here $g \in G$,  $A_\pm\in \mf{h}$=alg$(H)$,  
$k$ is a  level  and $m$ is a parameter defining the mass 
of  elementary excitations near $g=1$.\foot{We use the following 
 notation. We choose Minkowski signature in 2 dimensions,  with 
 $d^2x=dx_0 dx_1$,  \  $\dpm\equiv \del_0 \pm \del_1$.
 $\ind$ is the index of the representation of 
$G$ in which $g$ is taken as  a matrix.
 For $G=SU(N)$,
$\ind=\fr{1}{2}$ for the fundamental representation and
$\ind = \dcn = N$ for the adjoint representation, where
$\dcn$ is the dual Coxeter number. For $G=SO(N)$ the values
are $\ind=1$ and $\ind=\dcn=N-2$ respectively.}
The standard  symmetric gauging corresponds to  $\tau=\id$
($\tau$ is an automorphism of  $\mf{h}$); 
for an  abelian gauge group $H$ there is an option of axial gauging 
corresponding to   $\tau(h)=-h$.
The constant matrix $T$ defining the potential is chosen to 
commute with  $H$ (see, e.g., 
 \ci{gt,mirp} for  details).


Recent interest in  such  models  is due to  their 
relation, via the Pohlmeyer 
reduction, to classical string theory on 
 symmetric spaces \ci{pol,devsan,mi1,
gt,gt1,mirp,gevic,hmsol}. 
In the case of the  \adss superstring theory  the classical Pohlmeyer 
reduction leads to  a special integrable  massive 2d 
 theory defined by the 
 $G/H=[SO(1,4) \times SO(5)]/[SO(4) \times SO(4) ]$
 gauged WZW 
model with  an integrable potential and coupled to a particular set of 
2d fermions
\ci{gt,ms}.  The corresponding quantum  theory has certain unique features 
(it is UV finite \ci{rt} and is closely related to the original 
\adss superstring  at the one-loop level \ci{hit,iwa}). This  suggests 
 \ci{gt} that it may, in fact, be  quantum-equivalent 
to the \adss superstring. 
If that were   indeed the case,   this 
theory could be used
as a starting point for a 2-d Lorentz covariant
``first-principles'' solution of the  $AdS_5 \x
S^5$  superstring  based on finding an  exact soliton   S-matrix, just as 
in the case of  standard   2d sigma models \ci{zz} or for some 
similar massive theories \ci{dh,hmsol} (see also \ci{dorr}).

An important  check of a 
proposal for  an exact   quantum  soliton   S-matrix of an integrable theory 
would be  to demonstrate its consistency  with the 
perturbative S-matrix computed from  the   path integral 
defined by  the classical action.  

This motivates the study of  perturbative  S-matrices of 
such generalized sine-Gordon models with a non-abelian gauge symmetry $H$.
In \ci{ht} we computed the tree-level two-particle S-matrix 
for the   bosonic $SO(N+1)/SO(N)$   theory 
and then  for  the full reduced theory associated to the \adss superstring. 
The resulting  S-matrix  exhibited some remarkable features, 
in particular,  it group-factorised in the same way as the
 non Lorentz invariant tree-level 
$\mf{psu}(2|2)\oplus \mf{psu}(2|2)$ 
invariant 
light-cone gauge S-matrix \ci{kmrz} of the 
 $AdS_5 \x S^5$ superstring.  It also  had an  intriguing 
similarity with the classical $\mf{psu}(2|2)$ 
 trigonometric  r-matrix 
of  \ci{bek,beisctr}.\foot{As discussed in \ci{beisctr}, 
in a certain limit the coefficients of this r-matrix match
those of the perturbative S-matrix of  \ci{ht}, up to some
constant matrix which breaks the explicit $[SU(2)]^2$
structure to its Cartan subgroup.}

The next important
 step towards  unravelling the full structure of the S-matrix of the 
 reduced theory 
 is to extend  the tree-level computation of \ci{ht} to the one-loop level.
In this paper we address this  problem 
  for the  bosonic $SO(N+1)/SO(N)$   theory, 
hoping to return to the complete 
 $[SO(1,4) \times SO(5)]/[SO(4) \times SO(4) ]$  theory
 with fermions 
   in the future.

 There are technical issues involved in computing the perturbative 
 S-matrix for the generalised sine-Gordon theories defined by  \rf{gwzwact}.
 Such models were  mostly  studied  for abelian gauge groups $H$
\ci{fpgm,mirpa,mirfp,cam,mirat}, of which the 
complex sine-Gordon model  is a prime  example. In this case there is 
an option of axial gauging, in which case  the vacuum is unique up to gauge
transformations. 
In the case of a non-abelian  $H$  there is a 
 non-trivial vacuum
moduli space, no global symmetry and on integrating out the
gauge fields $A_\pm$  one is left with a Lagrangian that has no
perturbative expansion about the trivial vacuum.
As was argued in \ci{ht}, the problem with 
expansion is an artifact of the gauge fixing procedure 
on $g$. If instead one chooses the ``light-cone'' gauge 
 $A_+=0$  \ci{ht} one is  able  to construct a
perturbative Lagrangian for the asymptotic excitations and
thus compute the tree-level S-matrix. 
Going beyond tree level requires, however,  
taking into account the  non-trivial  contribution 
of the delta-function   constraint resulting from integrating 
 over $A_-$ in the  path integral for  the action \rf{gwzwact}.

\



Below we will  start with the 
case of the $SU(2)/U(1)$ axially gauged theory 
corresponding to the  complex sine-Gordon (CsG) model.
The tree-level S-matrix of this theory   \ci{dvm1} satisfies the
 Yang-Baxter equation. The  Yang-Baxter equation,   however, is violated 
at the one-loop level for the ``naive''
S-matrix  obtained directly from the  standard CsG
action, following from \rf{gwzwact} upon solving for $A_\pm$. 
The  quantum theory based on this  CsG action   was studied in 
 \ci{dvm1,dvm2,maillet} where  it was suggested to add a quantum
counterterm  to restore  factorised scattering.
 In reference \ci{dh} the exact  quantum soliton  S-matrix satisfying the Yang-Baxter 
 equation was proposed. It
was conjectured that if  the quantum  CsG model is defined 
in terms of the
 $SU(2)/U(1)$ gauged WZW theory   \rf{gwzwact}
then  the necessary quantum counterterms 
required to obtain a  factorizable perturbative S-matrix consistent 
with the exact S-matrix should  appear automatically.

We will demonstrate that this is indeed the case at the one-loop 
level. We will show that equivalent ``counterterms''
(leading to the same S-matrix)   can be obtained  (i) from 
 the quantum effective action of the gauged WZW model \ci{t1}, 
or (ii)  directly from the determinant resulting from integrating out 
the non-dynamical fields  $(A_+, A_-)$, or (iii) from  the delta-function constraint 
in the $A_+=0$ gauge. 

We will then extend the $A_+=0$ gauge approach to the general 
 $SO(N+1)/SO(N)$ theory. 
 In the  case of the non-abelian gauge group $H=SO(N)$ 
 the   resulting 
 S-matrix will be  found to violate the  standard Yang-Baxter  equation
 already at the   tree level, but we will suggest   that this should 
  not contradict the quantum integrability
 of the theory  being  a   gauge  artifact 
  (the excitations we scatter transform non-trivially  under 
 unbroken global part of the gauge group). 
 
 Further clarification of this 
  issue  and  the extension to the full reduced theory for the \adss superstring 
 are  among the important problems for the future.

\



The structure of the rest of this paper is as follows.
In section 2 we review the $A_+=0$ gauge
approach  used to compute the
tree-level S-matrix in \ci{ht}. We show that fixing this gauge  and 
integrating out the unphysical degrees of freedom gives rise
to a  functional determinant in the path integral, which leads to 
  a non-trivial 
contribution  to the one-loop S-matrix.

In section 3 we discuss the case of the  $SU(2)/U(1)$
theory  related to  the 
complex sine-Gordon model. We compute the  corresponding 
one-loop S-matrix
using three different methods, one of which uses  the $A_+=0$ gauge.
  They  all give the same  result,  equal to the sum  of 
  the ``naive'' one-loop
S-matrix of \ci{dvm1}  with  a  non-trivial correction.
The total   one-loop S-matrix  matches the expansion of
the (charge-one sector of)  the quantum soliton S-matrix of \ci{dh}.

In section 4 we generalise the computation 
of the one-loop S-matrix in the  $A_+= 0$ gauge 
to the non-abelian  $G/H=SO(N+1)/SO(N)$  theory. 
The resulting S-matrix  turns out 
to violate the standard Yang-Baxter equation 
in the case of  non-abelian gauge symmetry.
We discuss the remarkably simple structure of this 
violation at the tree level and  possible ways 
to resolve  an apparent contradiction with the expected 
quantum integrability of the underlying theory.


The appendix contains a demonstration of a  relation   between  two 
effective Lagrangians used to compute the one-loop complex
sine-Gordon  S-matrix in section 3.

\renewcommand{\theequation}{2.\arabic{equation}}
\setcounter{equation}{0}

\section{Gauged WZW theory with  an integrable potential
      \\  in the gauge $A_+=0$\la{rev}}

In this section we shall start by reviewing the result  of
\ci{ht} for the tree-level S-matrix of the massive excitations of the 
 gauged WZW action with an 
integrable potential, \eqref{gwzwact}. We will then
generalise the computation of the S-matrix  to the one-loop order. 

As in \ci{ht} we assume that $G$ is a compact group and that
the corresponding algebra $\mf{g}$ admits an orthogonal
decomposition
\be\la{ordec}
\mf{g}=\mf{h}\oplus\mf{m}\,,
\ee
where $\mf{h}$ is the algebra corresponding to the gauge
group $H$, and $\mf{m}$ is the orthogonal complement of
$\mf{h}$ in $\mf{g}$. We also assume that $G/H$ is a
symmetric coset space, i.e. 
\be\la{commu}\ba{ccc}
\com{\mf{h}}{\mf{h}}\subset \mf{h}\,,
&\com{\mf{h}}{\mf{m}}\subset \mf{m}\,,
&\com{\mf{m}}{\mf{m}}\subset \mf{h}\,,
\\&\Tr(\mf{m}\,\mf{h})=0\,,&
\ea\ee
and that we have an orthogonal basis of a matrix
representation of $\mf{g}$, the elements of which we denote
as follows\la{alg}
\bit
\item $\TT_i$, where $i=\dim\mf{m}+1,\ldots,\,\dim\mf{g}$ \ \ is a 
      basis for $\mf{h}$;
\item $\TT_a$, where $a=1,\ldots,\, \dim\mf{m}$ \ \ is a basis for 
      $\mf{m}$;
\item $\TT_A=\{\TT_a,\TT_i\}$, where $A=1,\ldots,\,\dim\mf{g}$ \ \ is 
      an orthogonal basis for $\mf{g}$, that is
      \be 
      \Tr(\TT_A \TT_B) = -K^2\de_{AB}\,. \la{noo}
      \ee
       \eit
$K$ is a constant that will be fixed so that the quadratic
kinetic terms in the action have canonical form.  $K$ 
will depend on  the  coupling $k$, in particular,  $K\sim k^{-\fr{1}{2}}$.

The structure constants of the algebra
$\mf{g}$ are defined by\foot{$G$ is compact, so  we will not distinguish between 
raised and lowered indices.}
\be
\com{\TT_A}{\TT_B} = f_{ABC}\TT_C\,,
\ee
with  $f_{ABC}$ being  completely antisymmetric. From the
commutation relations \eqref{commu} we have
$f_{aij}=f_{abc}=0$. Due to the  normalization of the
generators in \eqref{noo} the
structure constants will also have the coupling dependence,
$f_{ABC} \sim k^{-\fr{1}{2}}$.

As in \ci{ht},  we shall fix  the gauge $A_+=0$ in the path integral defined by 
 the action $\mc{S}[g,A_+,A_-]$ in  \eqref{gwzwact} and integrate
out $A_-$, getting 
\be\la{pint}
\mc{Z} = \int [ d g] \;\;\delta\big[(g^{-1}\dpl g)|_{\mf{h}}\big]\;\;
                   e^{i\mc{S}[g,0,0]}\,.
\ee
The gauge   $A_+=0$
  preserves the   Lorentz invariance in 2 dimensions   but
like  the analogous light-cone  gauge 
 in Yang-Mills  theory it  does not fix the global part of 
the gauge group  $H$. 


\subsection{Expansion of the gauge-fixed action\la{revact}}

To compute the S-matrix we will need to expand the action,
\eqref{gwzwact}, near a trivial 
 vacuum  $g=1$ and  solve the delta-function constraint in
\eqref{pint}.
 Following \ci{ht}, we
parametrise $g$ as follows
\be\la{tak}
g=e^{\h}\,, \hs{30pt} \h\in \mf{g}\,,
\ee 
and expand  in $\h$,  \foot{$\mf{L}_\h $ is the
usual Lie derivative defined as\    $ \mf{L}_\h(\zeta) =
\com{\h}{\zeta}\,.$}$^{,}$\foot{We should also note
that the potential terms in \eqref{gwzwex1} containing an
odd number of $\h$ fields vanish due to the relation,
$\Tr(\mf{L}_\h (A)\; B) = -\Tr( A\;\mf{L}_\h (B))$.}
\be\la{gwzwex1}
\!\!\,\,\,
\mc{S}[g(\h),0,0] = - \fr{1}{8\pi\ind}\ \int d^2x \; \ 
  \trm{Tr} \Big(
  \sum_{n=0}^{\infty} \fr{k}{(n+2)!}
     \big[\;\dpl \h\, \mf{L}^{\;n}_\h (\dm \h)
        - \gok \;\mf{L}_\h(T)
              \,\mf{L}^{\;n+1}(T)\big]\Big)\,.
\ee
Using the orthogonal decomposition of $\mf{g}$,
\eqref{ordec}, we split $\h$  as  
\be\la{etaxxi}
\h = X + \xi\,, \hs{30pt} X \in \mf{m}\,, 
                \hs{10pt}\xi\in \mf{h}\,.
\ee
The coset component $X$ represents  $\dim\,\mf{m}$ 
massive  asymptotic excitations (as seen by solving the
gauge-fixed  linearised equations of
motion), for which the tree-level two-particle S-matrix was
found in \ci{ht}.  Here  we will  compute the
one-loop correction to their   perturbative S-matrix. 

At the classical level we can use the $\delta$-function 
constraint in \eqref{pint}  to solve for the ``gauge
component''  $\xi$ in terms of  $X$. 
Expanding $\big(g^{-1}\dpl g\big)\big|_{\mf{h}}=0$ to cubic
order in fields we get
\be\la{conex}\begin{split}
\dpl \xi & - \fr{1}{2}\com{X}{\dpl X}
           - \fr{1}{2}\com{\xi}{\dpl \xi}
      \\ & + \fr{1}{6}\com{X}{\com{X}{\dpl \xi}}
           + \fr{1}{6}\com{X}{\com{\xi}{\dpl X}}
           + \fr{1}{6}\com{\xi}{\com{X}{\dpl X}}
           + \fr{1}{6}\com{\xi}{\com{\xi}{\dpl \xi}}
           + \ldots = 0\,.
\end{split}\ee
Solving perturbatively for $\xi$   gives
\be\la{xi0x}
\xi_0 [X] = \fr{1}{2}\fr{1}{\dpl}
                    \com{X}{\dpl X}+\ord{X^4}\,.
\ee
As $\xi_0[X] \sim {X^2}$,
when $\xi=\xi_0[X]$ is substituted into the action there
will be no cubic terms in $X$. Thus, to compute the one-loop
S-matrix,  we only need  the part of the  action which is of 
quartic order in 
$X$. Substituting \eqref{etaxxi} and \eqref{xi0x} into 
\eqref{gwzwex1}, using integration by parts and the
Jacobi identity, we end up with the following  remarkably simple 
local action  \ci{ht}, 
\bea\la{gwzwex3}
\mc{S}[g(\h(X,\xi_0[X])),0,0] 
        &\!\!\! = &\!\!\! - \fr{k}{8\pi\ind} \int  d^2x \; 
            \trm{Tr} \Big[\,
            \fr{1}{2} \dpl X \dm X 
          - \fr{\gok}{2} \com{X}{T}\com{X}{T}
\\& & \hs{15pt}+\  \fr{1}{12}\com{X}{\dpl X}\com{X}{\dm X}\no
      + \fr{\gok}{24}\com{X}{\com{X}{T}}\com{X}{\com{X}{T}}
      + \ord{X^6}\Big]\,.\no
\eea
This action can be used to compute  the tree-level
two-particle S-matrix \ci{ht}, as well as the  part of  the
one-loop two-particle S-matrix given by  bubble
diagrams.
The only additional piece of information one needs to know
is the commutation relations with the  matrix  $T$ in the potential,
which 
depend on the particular choice of the groups $G$ and $H$.
The 
calculations then reduce to standard Feynman integrals.

Let us note that the renormalization of the general theory \rf{gwzwact} was discussed 
in \ci{rt}. The WZW coupling $k$ is not, of course,  renormalized but 
there is a logarithmic  UV renormalization of the mass parameter $m^2$
(as well as  a field renormalization).  One may choose a scheme 
($\overline{\rm MS}$ scheme) in which there is no finite  renormalization 
of $m^2$  at the one-loop order. This will be assumed below.


The aim of the rest of this  section  will be   to discuss the non-trivial contribution of the 
path integral constraint in \rf{pint} 
to the one-loop
S-matrix. Thus we will not present the detailed expressions
for the S-matrix terms arising from the Lagrangian. The
explicit results will be given for the particular  cases of
$G/H=SU(2)/U(1)$ and $G/H=SO(N+1)/SO(N)$ in later sections.

\subsection{Functional determinant contribution\la{revfd}}

 At the quantum level, solving the
delta-function constraint \eqref{conex}    to eliminate
$\xi$   from  the path integral \eqref{pint}   will give
rise to a field-dependent functional
determinant.
To find it we  functionally 
differentiate the constraint equation \eqref{conex} and
evalute the resulting operator on $\xi = \xi_0[X]$ in 
\eqref{xi0x}. The resulting   contribution to the path integral   is then 
 given by
\be \la{op}
\big( \det \mQ \big)^{-1} \ ,
\ee
\bea
&& \mQ  \zeta   = \dpl \zeta
  - \fr{1}{4}\com{\fr{1}{\dpl}
             \com{X}{\dpl X}}{\dpl \zeta}
  + \fr{1}{4}\com{\com{X}{\dpl X}}{\zeta} \cr 
&& \hs{50pt}+ \  \fr{1}{6}\com{X}{\com{X}{\dpl \zeta}}
     - \fr{1}{6}\com{X}{\com{\dpl X}{\zeta}}
     - \fr{1}{6}\com{\com{X}{\dpl X}}{\zeta}
     + \ord{X^4}
\,, \la{qu}
\eea
where the operator $\mQ$ acts on 
a function $\zeta$  taking values in $ \mf{h}$.


Let us   rewrite the operator $\mQ$  in the following form 
\be\la{opd}
\mQ= \vec{\dpl}(1+\al[X]) 
       + \bet[X]   \,,
\ee
where the $ \;\, \vec{} \;\, $ symbol denotes the operator
 acting all the way to the right. $\al$ and $\bet$ are
 functions of $X$ and $\del X$  that are both $\ord{X^2}$.\foot{For
notational ease we have suppressed a Lorentz $_+$ index on
$\bet$. We will denote the $\ord{X^n}$ part of $\al$ and
$\bet$ as $\al_n$ and $\bet_{n}$ respectively. Note that due
to the group structure $n$  takes  only even values.}

The prescription we shall  use is to expand the determinant
of $\mQ$ 
\eqref{opd} in the usual perturbative way, treating
$\vec{\dpl}$ as the free part. We shall  ignore all
quadratic divergences and  tadpole contributions. 
Then  to evaluate the contribution of the determinant to the
two-particle one-loop S-matrix  we may 
ignore $\al [X]$ and $\bet_{n}[X]$ for $n\geq4$.
Factorising out the free part of the
operator, we have 
\be\la{trdet}
\ln (\det  [\dpl^{-1}\mQ])^{-1}  = -
\tr \ln [\dpl^{-1}\mQ]\,.
\ee
One may then extract the $\ord{X^4}$ part,
\be\la{above}
- \tr\big(\al_4[X]+\dpl^{-1}\bet_{4}[X]\big) 
- \fr{1}{2} \tr \big(\al_2[X]+\dpl^{-1}\bet_{2}[X]\big)^2\,.
\ee
The traces of $\al_4[X]$ in the first term and of 
$\al_2[X]^2$ in the second term both give quadratic
divergences, which we ignore. The trace of 
$\dpl^{-1}\bet_{4}[X]$ in the first term and the cross-terms
in the second term both give tadpole integrals which we also set to zero. 
We are then  left with 
\be
- \tr \big(\dpl^{-1}\bet_{2}[X]\big)^2\,
\ee
from the second term in \eqref{above}. 


\
Let us note that moving the free operator  $\vec{\dpl}$   all the way
to the left and setting tadpoles to vanish can be
reinterpreted as choosing a different parametrisation of
$g$.
For  example, we may  choose instead of \rf{tak},\rf{etaxxi}  the 
following parametrisation\foot{Here $\al_2[X]$ is understood 
as an operator acting on $\xi$ by commutators.
} 
\be\la{takk}
g=e^\h\,, \hs{20pt} \h = X + \xi - \al_2[X] \xi\,.
\ee
The operator  \eqref{opd} would be corrected 
 to $\ord{X^2}$ as follows
\be\la{opds}
\mQ= \vec{\dpl}(1+\al[X]) 
       + \bet[X]  -\vec{\dpl}\al_2[X]+\ord{X^4}
       =  \vec{\dpl}
       + \bet_2[X]  +\ord{X^4} \,,
\ee
i.e. the $\ord{X^2}$
part of $\al[X]$  will cancel.\foot{It does not seem to be 
possible to do a similar change of parametrization 
to cancel the  $\ord{X^2}$  part of  $\bet[X]$ without redefining
the gauge field $A_-$.}
Thus the prescription detailed above 
can be seen to be equivalent to  a field redefinition of $\xi$ by
some function of $X$ and $\del X$   or   choosing an
alternative measure for  $\xi$ in the path integral.\foot{Also, having a different
ordering    of the 
 operator $\vec{\dpl}$  with respect to $X$-dependent factors
 in $\mQ$ could be interpreted as 
choosing an alternative measure on $A_-$.} 

As we shall see  below, in the
complex sine-Gordon case computing the determinant contribution 
with   this prescription will  give  corrections
to the S-matrix that match the results found using  two alternative methods 
described in
sections \ref{v1} and \ref{v2} and also match  the soliton S-matrix
of \ci{dh}. 

Next, 
let us  expand  $X$  in generators 
\be
X=X_a\TT_a\,, \la{cx}
\ee
and use the Jacobi identity to rewrite the determinant   \eqref{op} of \eqref{opd} in the
form 
\be\la{op3}\begin{split}
-\tr \ln  \bigg[\vec{\dpl} & 
          \Big(\de_{ij} 
        - \fr{1}{4}\fr{1}{\dpl}(X_a \dpl X_b)
                                         f_{abk}f_{ijk}
        - \fr{1}{6}X_a X_b f_{aci}f_{bcj}\Big)
\\ & \hs{100pt}
        + \fr{1}{2}X_a \dpl X_b f_{aic}f_{bjc}
        + \ord{X^{-4}}\bigg]\,.
\end{split}\ee
 Following
the above  prescription,  let us  now ignore the term corresponding to
$\al[X]$ in \eqref{opd},
getting
\be\la{op4}\begin{split}
-\tr \ln \big(\de_{ij} \dpl + \fr{1}{2}  V_{abij}     X_a \dpl X_b
                           \big)\,, \ \ \ \ \ \ \ \ \ \ \ \ \ \ 
 V_{abij} \equiv  f_{aic}f_{bjc} \ . 
\end{split}\ee
We may then follow the standard perturbative
approach\foot{Recall that the  coupling $k^{-1}  \ll 1 $ 
dependence is  contained in the structure constants.} using
the Feynman rules in figure \ref{fig2a}, 
that come from \eqref{op4} to compute the Feynman diagrams
in figure \ref{fig2}.

\begin{figure}[t]
\begin{center}
\epsfig{file=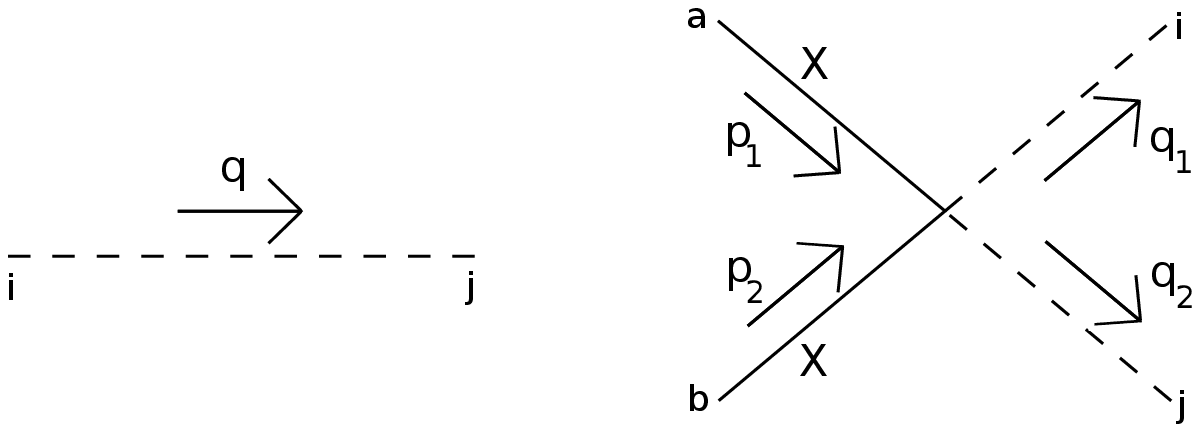, width=13cm}
\vs{-0.5cm}
\be\hs{110pt}\fr{i\de_{ij}}{q_+}\,, \hs{95pt} 
\fr{i}{2}\Big[(p_{1+}+p_{2+})V_{(ab)ij}
         -(p_{1+}-p_{2+})V_{[ab]ij}\Big]\,.\no
\ee
\end{center}
\caption{Feynman rules\la{fig2a}}
\end{figure}

\begin{figure}[t]
\begin{center}
\epsfig{file=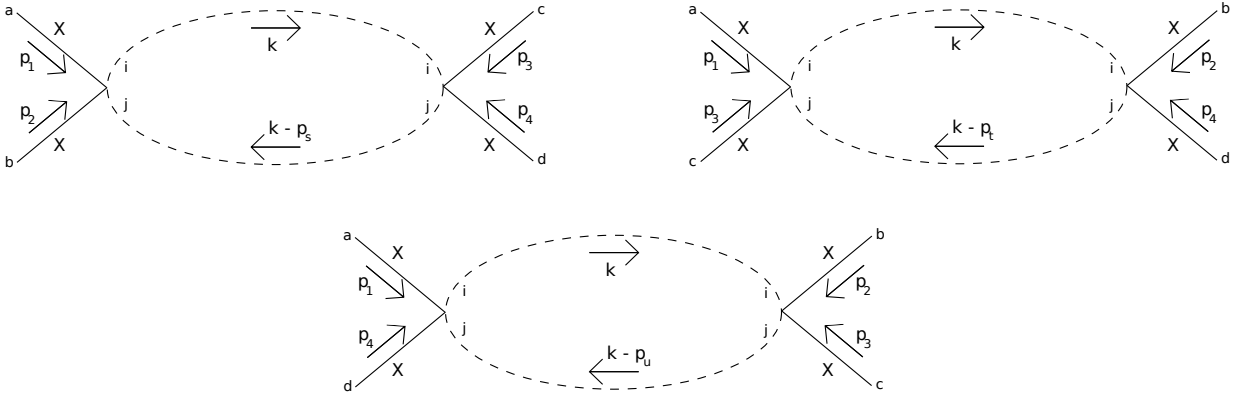, width=16.5cm}
\end{center}
\caption{One-loop  Feynman diagrams \la{fig2}}
\end{figure}

After solving the 
momentum conservation constraint and the on-shell condition with
\be
p_1=-p_3=m e^{\q_1}\,,\hs{30pt}p_2=-p_4=m e^{\q_2}\,, \la{li}
\ee
we have
\be\ba{cc}
\q \equiv \q_1 - \q_2 \ , 
\\-p_s^2=p_{s+}p_{s-}=4m^2\cosh^2\fr{\q}{2}\,,  
\ \ \ \ \ \ \ \ \ p_s=p_1+p_2\,,
\\ \ \ \ \ \ \ -p_t^2=p_{t+}p_{t-}=0\,, 
\ \ \ \ \ \ \ \ \ \ \ \ \ \ \ \ \ p_t=p_1+p_3,
\\-p_u^2=p_{u+}p_{u-}=4m^2\sinh^2\fr{\q}{2}\,, 
\ \ \ \ \ \ \ \ \ p_u=p_1+p_4 \,.\la{lii}
\ea\ee
Given that 
\be 
V_{(ab)ij}V_{[cd]ji} = 0\,,
\ee
and using the standard integral 
 (see, e.g., \ci{poly})\foot{This integral can be done by 
using the Lorentz-covariant 
 prescription $\fr{1}{k_+} \to  \fr{k_-}{k^2 + i \epsilon} 
=  \fr{1}{k_+  + i \epsilon \sign (k_+)} $  (where $k$ stands for $q$ or $q-p$) 
 and integrating 
separately over $q_+$ and
$q_-$  as in \ci{poly}.
An an equivalent result is found 
by  using  $\fr{1}{k_+} \to  \fr{ k_-}{k^2}$  and 
doing the integral directly by 
symmetric integration.}
\be
\int \fr{d^2 q}{(2\pi)^2} \; \fr{1}{q_+(q_+-p_+)}
   = \fr{i}{4\pi}\fr{p_-}{p_+}\ ,
\ee
the contributions of the $s,t,u$-channel diagrams in figure 
\ref{fig2} to the one-loop S-matrix are found to be 
respectively
\be\la{nabcont}\begin{split}
\Delta S_1=  -\fr{i}{16\pi\sinh\q} &
   \Big(\cosh^2\fr{\q}{2} \ V_{(ab)ij}V_{(cd)ji}
        +\sinh^2\fr{\q}{2}\ V_{[ab]ij}V_{[cd]ji}\Big)\,,
\\\Delta S_2=-\fr{i}{16\pi\sinh\q} &
   \Big(\cosh\q \  V_{[ac]ij}V_{[bd]ji}\Big)\,,
\\ \Delta S_3=\hs{8pt}\fr{i}{16\pi\sinh\q} &
   \Big(\sinh^2\fr{\q}{2}\ V_{(ad)ij}V_{(bc)ji}
        -\cosh^2\fr{\q}{2}\ V_{[ad]ij}V_{[bc]ji}\Big)\,.
\end{split}\ee
 Here we have included the usual Jacobian
factor  $\fr{1}{4m^2\sinh\q}$ arising from solving the
momentum conservation constraint.

Note that the  second of these contributions is somewhat ambiguous when
taking the limit $p_t \ra 0$.
 Consider the integral
\be\la{sample}\begin{split}
&(p_{1+}-p_{3+})(p_{2+}-p_{4+})
\int \fr{d^2q}{(2\pi)^2}\fr{1}{q_+(q_+-p_{t+})}
=\fr{i}{4\pi}(p_{1+}-p_{3+})(p_{2+}-p_{4+})
     \fr{p_{t-}}{p_{t+}}
\\=&-\fr{i}{4\pi}(p_{1+}-p_{3+})(p_{4+}-p_{2+})
    \bigg(\fr{p_{1-}+p_{3-}}{p_{1+}+p_{3+}}\bigg)^x
    \bigg(\fr{p_{2-}+p_{4-}}{p_{2+}+p_{4+}}\bigg)^{1-x}\,,
\end{split}\ee
where we have used that  $p_t=p_1+p_3 = -(p_2+p_4)$  and introduced an arbitrary parameter $x$. Substituting
in for the on-shell momenta in terms of the  rapidities
$
p_{i\pm}=m e^{\pm\q_i}\,,
$
and taking the limits
$
\q_3 \ra i\pi+\q_1, \ \ \q_4 \ra i\pi + \q_2\,,
$
to solve $p_t=0$, we find  that the integral  \eqref{sample} 
is
\be\la{sampterm}\begin{split}
&\fr{i m^2}{\pi}e^{(1-2x)\q}\,,
\end{split}\ee
and thus depends on the arbitrary parameter $x$. 
 We may  fix this ambiguity by 
demanding that the resulting S-matrix should satisfy the
physical requirements of crossing symmetry and unitarity.
Noting that the integral in \rf{sample} appears in the S-matrix with the factor of 
 $V_{[ac]ij}V_{[bd]ji}$ it is clear that we
should take the average of two terms of the type
\eqref{sampterm}, one with $x=1$ and one with $x=-1$. This
then gives a consistent expression   proportional to $\cosh
\q$ presented in \eqref{nabcont}.

\renewcommand{\theequation}{3.\arabic{equation}}
\setcounter{equation}{0}

\section{Complex sine-Gordon model \la{csgdvm}}

In this section we will review some relevant aspects of the
complex sine-Gordon model  (CsG),  which at the classical level 
may be  defined as a special 2-d sigma model
with  a particular integrable   potential containing  two real 
bosonic  scalar fields  and having  global $SO(2)$ symmetry \ ($ a,b,= 1,2 $)
\be\la{lagcsg}
\Lag = \fr{1}{2}\Big(\fr{\dpl\psi_a\dm\psi_a}
                          {1-\zg^2\psi_b\psi_b}
                       - m^2 \psi_a\psi_a\Big)\,, 
\ee
where $\zg$ is a  dimensionless coupling and $m$ is the free mass
of the elementary excitations.

 We shall first   recall the perturbative
analysis of \ci{dvm1}  where it was  noticed that the scattering 
  factorisation property   is
broken at one loop, but may be  restored by the addition of  quantum
counterterms. We shall  then discuss the soliton S-matrix of
\ci{dh}, and its relation to the scattering of elementary 
excitations. 
In \ci{dh} it was suggested  that the theory that is 
quantum-integrable is actually   the   $SU(2)/U(1)$   gauged WZW model 
 with  an integrable potential \rf{gwzwact}. This theory,  which  reduces to the  
complex sine-Gordon action of \ci{dvm1}   at the classical level, 
 should then  be viewed as the  proper quantum definition of 
the latter.

We shall then  show how to construct the  one-loop counterterms required to 
preserve the  integrability directly from this   gauged WZW model, thus providing
  convincing   evidence for  the correctness of the
proposal of  \ci{dh}.

\subsection{Perturbative  S-matrix}

Following \ci{dvm1} we shall   compute the perturbative  S-matrix 
 of  CsG  model,
splitting  the Lagrangian \rf{lagcsg} into  free  and 
interacting
parts
\be\la{lagcsgex}
\Lag = - \fr{1}{2} \del_+\psi_a\del_-\psi_a 
       - \fr{m^2}{2} \psi_a\psi_a 
       + \fr{\zg^2}{2} \psi_b\psi_b 
                     \del_+\psi_a\del_-\psi_a 
       + \ord{\zg^4}\,.
\ee
 The S-matrix can
be written terms of the  three functions $S_1(\q)$, $S_2(\q)$, 
$S_3(\q)$, 
\be\la{smatso2}
\bra{\psi_c(-p_3)\psi_d(-p_4)}\mbb{S}         
\ket{\psi_a(p_1)\psi_b(p_2)} = (2\pi)^24\ep_{p_1}\ep_{p_2}
\de(p_1+p_3)\de(p_2+p_4)\ S_{ab,cd}(\q)\,,
\ee
\be
S_{ab,cd}=S_1(\q)\ \de_{ab}\de_{cd} 
+\ S_2(\q)\ \de_{ac}\de_{bd}+ \ S_3(\q) \
\de_{ad}\de_{bc}\,, \la{sesd} 
\ee
where $\mbb{S}$ is the usual scattering operator and 
$\ep_p=\sqrt{p^2+m^2}$ is the on-shell energy associated 
with the  spatial momentum  $p$. $\q$ is the difference of
the rapidities of the two  on-shell massive particles. The
S-matrix is restricted to take this form by
the Lorentz and global $SO(2)$ symmetries. Crossing symmetry
also implies the following relations, 
\be\la{cross}
S_1(\q)=S_3(i\pi-\q)\,,\hs{30pt}S_2(\q)=S_2(i\pi-\q)\,.
\ee
In \ci{dvm1} this S-matrix was computed to one-loop order with the result
\be\la{smatcsg}\begin{split}
S_1(\q) = & - i\zg^2 \coth \q 
            + i \fr{\zg^4}{2\pi}(\cosech \q + \coth \q) 
            + \zg^4 \coth \q\ \cosech \q +\ord{\zg^6}\,,
\\S_2(\q) = & 1 + i \zg^2 \cosech \q 
                - \zg^4 (\fr{1}{2}+\cosech^2\q) +\ord{\zg^6}\,,
\\S_3(\q) = & i \zg^2 \coth \q 
              + i \fr{\zg^4}{2\pi}(\cosech \q - \coth \q) 
              - \zg^4 \coth \q\ \cosech \q +\ord{\zg^6}\,.
\end{split}\ee

Factorised scattering and thus integrability of 
 a 2d theory implies,  in general, that the  Yang-Baxter equation 
is  satisfied.
  In the 
 abelian case like the CsG   theory of
an $SO(2)$ doublet this  condition 
can be reduced to the reflectionless scattering condition, i.e.  to  the
 vanishing of the reflection  coefficient 
\be\la{r}
R(\q)\equiv S_1(\q)+S_3(\q)=0\,.
\ee
We can immediately see that while  this property  is true at the  tree  level 
(i.e. to order $\zg^2$), it  breaks down at  
one-loop order:   we find from \rf{smatcsg} 
\be\la{rcsg}
R(\q)=\fr{i \zg^4}{\pi}\cosech\q+\ord{\zg^6}\,.
\ee
In \ci{dvm1}  it was suggested to add  a quantum counterterm to
restore the factorised scattering at one-loop order.
 If such   counterterm is    required
to  be ultra-local (no derivatives) then it is found  to be   unique
\be\la{dvmcounter}
\De \Lag=\fr{m^2 \zg^4}{4\pi}\psi_a\psi_a \psi_b\psi_b\,.
\ee
Adding its contribution  leads to  the following
``corrected''  S-matrix 
\be
\hS = S + \Delta S\,, \la{joi}
\ee
\be\la{smatcsgsh}\begin{split}
\hS_1(\q)= & - i \zg^2 \coth \q 
                  + i \fr{\zg^4}{2\pi}\coth \q 
                  + \zg^4 \coth \q\ \cosech \q +\ord{\zg^6}\,,
\\  \hS_2(\q) = & 1 + i\zg^2 \cosech \q 
                       - i \fr{\zg^4}{2\pi}\cosech \q
                       -\zg^4 (\fr{1}{2}+\cosech^2\q) 
                       +\ord{\zg^6}\,,
\\   \hS_3(\q) = & i \zg^2 \coth \q 
                     - i \fr{\zg^4}{2\pi} \coth \q 
                     - \zg^4 \coth \q\  \cosech \q 
                     + \ord{\zg^6}\,, 
\end{split}\ee
satisfying the factorisation (or reflectionless \rf{r}) property.
The authors of   \ci{dvm1,dvm2}  conjectured that such  counterterm addition 
procedure   should  apply  to all orders, 
leading to a  factorizable quantum S-matrix. 

Let us mention that  while in \ci{dvm1} the counterterm was
restricted to be
ultra-local,   this is not, in fact,   a necessary requirement:  
the only condition is that the resulting S-matrix 
 should be reflectionless (or  satisfy the  Yang-Baxter equation)
   at the quantum level. 
For example, one may consider  local  counterterms 
with up to  two derivatives that 
may be interpreted as 
corrections to the sigma model part of the action.
 This  will be  relevant below in the context  of
the  gauged
WZW theory  interpretation of this model.

\

As implied by  the name of the   complex
sine-Gordon model, it 
can be  interpreted  as a theory for a single complex 
scalar field $Z$  
\be\la{cc}
Z=\fr{1}{\sqrt{2}}(\psi_1+i \psi_2)\,.
\ee
Using the global $U(1)$ and the 2d   Lorentz symmetry
 the   scattering matrix 
 may  be  represented  in the complex $Z$ basis as 
\be\la{smatu1}\begin{split}
\mbb{S}\ket{Z(p_1)Z(p_2)} = & 
                  S_{++++}(\q)\ket{Z(p_1)Z(p_2)}\,,
\\\mbb{S}\ket{Z(p_1)Z^{*}(p_2)} = & 
                  S_{+-+-}(\q)\ket{Z(p_1)Z^{*}(p_2)}
                + S_{+--+}(\q)\ket{Z^{*}(p_1)Z_(p_2)}\,,
\\\mbb{S}\ket{Z^{*}(p_1)Z^{*}(p_2)} = & 
                  S_{----}(\q)\ket{Z^{*}(p_1)Z^{*}(p_2)}\,,
\end{split}
\ee
where in  terms of the functions in \eqref{sesd}  we have 
\be\la{spm}\ba{c}
S_{++++}(\q) = S_{----}(\q) = S_2(\q) + S_3(\q)\,,
\\S_{+-+-}(\q) = S_1(\q) + S_2(\q)\,,
\hs{20pt}S_{+--+}(\q) = S_1(\q) + S_3(\q)\,.
\ea\ee
The reflectionless  scattering requirement \eqref{r}  then   
 implies that  
$S_{+--+}(\q)$ should vanish. If  this is the
case  and the   S-matrix has  the  crossing symmetry \eqref{cross}, then  it 
can be encoded in a single function, 
\be\la{tdsof}
S(\q)\equiv S_{++++}(\q)= S_2(\q) + S_3(\q) \,.
\ee
For the ``corrected''  S-matrix  \rf{joi},\eqref{smatcsgsh}  of \ci{dvm1}  we
then  get 
\be\la{pertct}
\hS(\q) = 1 + i ( \zg^2 
               -  \fr{\zg^4}{2\pi}) 
\coth\fr{\q}{2}
               -   \fr{\zg^4}{2}\coth^2\fr{\q}{2}
               + \ord{\zg^6}\,.
\ee

\subsection{Quantum soliton  S-matrix   in complex 
            sine-Gordon model \la{csgsol}}

Reference \ci{dh} considered the scattering of
non-topological 
solitons in the CsG model  and used the
semi-classical results of  \ci{dvm2}  and the requirement of the 
Yang-Baxter equation  to propose   the   full quantum
soliton  S-matrix for   this theory. As the  solitons of this
model  are not topologically distinct from the elementary 
excitations  considered above,   their S-matrices should be
related \ci{dvm2,dh}.

 To see this  let us  recall  that the
semiclassical mass spectrum of the charged
solitonic  states \ci{dvm2}  is given by
\be\la{solmasssc}
M(Q) = \fr{2m}{\zg^2}
\bigg|\sin\fr{\zg^2 Q}{2}\bigg|\,.
\ee
Here  $Q = \pm 1,\,\pm 2,\ldots,\,\pm Q_{max} $= 
$\big[\fr{\pi}{\zg^2}\big]$
is the  quantized $U(1)$ charge
of the soliton and $m$ is the free mass  of the elementary (or ``fundamental'') 
 excitation.\foot{Note 
that 
$\zg^2=\lambda^2/2$, where $\lambda$ is the coupling
used in \ci{dh}.} 
 Note that for the lowest charges, $Q=\pm
1$, the  soliton mass is $M(1)=m+\ord{\zg^4}$.
 Since the
 elementary fields    ($Z$ and $Z^*$ in \eqref{cc})  
have   charges $\pm 1$ and free mass $m$, one may identify  \ci{dvm2,dh} 
  the $Q=\pm 1$ solitons  with 
 the 
elementary excitations of the theory (referred to as 
``elementary mesons'' in \ci{dh}).\foot{
The expression \rf{solmasssc} is thus   consistent 
with the scheme choice  in which there is no finite renormalization of the mass
$m$ at the one-loop order.}

In \ci{dvm2}, using the counterterm \eqref{dvmcounter} of \ci{dvm1},
the one-loop correction to the
semiclassical mass  \eqref{solmasssc} was 
computed, giving the ``renormalized'' mass
\be\la{solmass}
M_r(Q) = \fr{2m}{\zg_r^2}
\bigg|\sin\fr{\zg_r^2 Q}{2}\bigg|\,,
\ee
where $\zg_r$ is the finitely renormalised coupling
\foot{From the results in \ci{dvm2} one can  see that if the
counterterm \eqref{dvmcounter} is not included, then the correction to the
soliton masses is of the same form as \eqref{solmass},
except that $\zg_r$  is given 
by 
$$\fr{2\pi}{\zg_r^{2}}  =\fr{2\pi}{ \zg^{2} } -  2  \ . $$\la{ftnt}} 
\be\la{gr}
\fr{2\pi}{\zg_r^{2}}  =\fr{2\pi}{ \zg^{2} } -  1   \   . 
\ee
This one-loop mass 
spectrum  \eqref{solmass}  was   conjectured to be exact. 

It was later   argued  in \ci{dh} that the only way to define
a consistent quantum soliton  S-matrix  is to require 
that the coupling $\zg_r$ in \rf{solmass} is quantized as\foot{We note again  that $\zg_r^2= \lambda_R^2/2$, where 
$\lambda_R$ is the finitely renormalised coupling used in 
\ci{dh}.}
\be\la{quant}
 \fr{2\pi}{\zg_r^2}=k  \in \mbb{N}\,, 
\ee
giving the following mass spectrum
\be\la{solmassk}
M_r(Q) = \fr{mk}{\pi}
\bigg|\sin\fr{\pi Q}{k}\bigg|\,.
\ee
The  soliton S-matrix  was then constructed  in \ci{dh} assuming 
factorised scattering;  it  can thus be written in terms of  one
function, \eqref{tdsof}. Extracting the full quantum S-matrix
for the elementary $(Q=\pm 1$) excitations from the general result
 of \ci{dh}
we get 
\be\la{smatsol}
\hS(\q) = 
\fr{\sinh\big(\fr{\q}{2}+i \fr{\pi}{k}\big)}
   {\sinh\big(\fr{\q}{2}-i \fr{\pi}{k}\big)}\,.
\ee
The pole of this S-matrix corresponds to the existence of
a $Q=2$ soliton that can form in the process of scattering of 
two $Q=1$ solitons. The location of the pole at $\q=\fr{2\pi
i}{k}$ follows  from the assumption that the one-loop mass
spectrum \eqref{solmassk} is exact.

Expanding \eqref{smatsol} for large $k$  (small $\zg_r$) gives
\be\la{smatsolex}
\hS(\q) = 1 + \fr{2\pi i}{k}\coth\fr{\q}{2} 
               - \fr{2\pi^2}{k^2}\coth^2\fr{\q}{2}
               + \ord{k^{-3}}\,.
\ee
This  matches \rf{pertct} provided 
$\fr{1}{k}= \fr{\zg^2}{2 \pi} - \fr{\zg^4}{(2 \pi)^2}  +  \ord{\zg^6}$
but this relation is not consistent with 
the identifications in \rf{gr} and \rf{quant}.
This problem may be attributed to the definition of couplings or  
 possible renormalization scheme freedom.



In \ci{dh} it was suggested  that this disagreement should  be
resolved if the quantum counterterms needed to preserve the integrability of the CsG 
model were understood as
arising from its definition in terms of the 
gauged WZW theory. 

\


The relation  between the complex sine-Gordon model  
and the $G/H= SU(2)/U(1)$  gauged WZW theory  with an
integrable potential was first discussed in \ci{bakas}.
 Classically,  if one fixes the $H$-gauge on $g\in G $  and
integrates out the gauge fields $(A_+,A_-)$ of the gauged
WZW theory then  the resulting Lagrangian is the CsG one  \eqref{lagcsg}. 

The relation to the  gauged WZW theory suggests an explaination of   the  
quantization condition on the 
coupling \eqref{quant} required by \ci{dh} for
consistency of the  quantum S-matrix for the solitons. 
Also, the integer shift in the relation 
between $\fr{2\pi}{\zg_r^{2}} $ and $ \fr{2\pi}{ \zg^{2} } $
in \rf{gr} (or in footnote \ref{ftnt}) is reminiscent of the quantum shift 
$k \to k+2$ in the $SU(2)$ WZW model. Details, however, depend  on the  precise 
definition of $\zg_r$ and the identification in \rf{quant}.
In particular, the presence of  quantum
counterterms like \eqref{dvmcounter} or   those arising from 
integrating out the $U(1)$  gauge fields of the gauged WZW theory, 
may  affect the finite coupling  renormalisation in a non-trivial way 
explaining possible   different  shifts in 
the relation between $\zg_r$ and $\zg$.


\

 In the remainder of this section we will 
 not refer to the 
 CsG Lagrangian \rf{lagcsg} directly, using the 
 gauged WZW theory \rf{gwzwact} 
 with the  coupling parameter $k$ as a starting point.
  We shall 
 investigate whether the perturbative expansion of this theory 
 is indeed in  agreement with the
mass spectrum \eqref{solmassk} and the  S-matrix \eqref{smatsolex}
of \ci{dh}.

\subsection{Gauged WZW   origin of  quantum     
            counterterms  \la{gwzw}}

Our  aim    will be  to understand the origin of the
quantum counterterms required  for maintaining   integrability 
 from the perspective of the  gauged WZW  formulation of
the CsG model.
The starting point will be  the action in   \eqref{gwzwact}
with  $G/H = SU(2)/U(1)$.
 We
define generators of $\mf{g}=\su(2)$,
\be\la{gen}
\ta_A=\fr{i}{2}\s_A\,,\ \ \ \ \ \  A=1,2,3 \ , 
\ee
where $\s_A$ are the usual Pauli matrices, and pick $\ta_3$
to be the generator of $H=U(1)$. The potential in  
\eqref{gwzwact} is then defined in terms of the matrix 
\be\la{ttt}
T = \ta_3\,. 
\ee
 $g$ will be  a matrix  in  the fundamental representation of
$G=SU(2)$ so that the  normalization constant in  \eqref{gwzwact} 
is    $\ind=\fr{1}{2}$. 

If we consider  the axially gauged case in \eqref{gwzwact} 
then   the automorphism of the algebra $\mf{h}=\mf{u}(1)$ 
is $\tau(a)=-a$.
 We can then  fix a gauge on $g$  as 
\be\la{param}
g=e^{-\ta_3 \chi}e^{2\ta_1 \phi}e^{\ta_3 \chi}\,,
\ee
where $\chi$ and $\phi$ are the  two remaining physical fields.
Substituting this  into the action 
\eqref{gwzwact} with 
 the  axial gauging choice
and then integrating out $A_\pm$ we end up with the
classical CsG  Lagrangian\foot{One can see
that this Lagrangian is equivalent to \eqref{lagcsg} by
using the following field and coupling 
redefinitions:
\be\ba{cc}\no
 \psi_1 = \sqrt{\fr{k}{2\pi}}\sin\phi\cos\chi\,, \ \ \ \ \ 
\psi_2 = \sqrt{\fr{k}{2\pi}}\sin\phi\sin\chi\,,
\ \ \ \ \ \ \ \ \zg^2 =\fr{2\pi}{k}\,.
\ea\ee
Thus  $\phi$ and $\chi$   may  be thought of as analogs of the  radial 
and angular coordinates  on the target space  with 
 $\psi_1$ and $\psi_2$ being analogs of the 
cartesian coordinates.}
\be\la{csgpc}
\mc{L} = \fr{k}{4\pi}
         \Big[  \dpl \phi \dm \phi 
                 + \tan^2\phi\ \dpl \chi \dm \chi 
            + \fr{\gok}{2}
              (\cos2\phi-1)\Big]\,.
\ee
While this  direct  integrating out of $A_\pm$  is  certainly 
valid   classically, there may be quantum 
corrections to the S-matrix resulting from  doing it consistently 
in the  path integral. We will study these corrections
to one-loop order in three different ways, all of which will  lead to  the same
result. The resulting S-matrix will  agree   
with (the $Q=\pm 1$ limit of) the soliton
scattering matrix  constructed in \ci{dh}.

The first  approach  will be  to start with  the quantum effective action 
for the gauged WZW theory proposed in  \ci{t1} 
and  deform it  by 
the  integrable potential  as in   \eqref{gwzwact}.
The  effective action of \ci{t1} is consistent with the 
quantum
conformal symmetry of the resulting sigma model.
 As the conformal
symmetry of the gauged WZW theory  is strongly related to the
integrability of the deformed theory, one may expect that the resulting 
 S-matrix will  have the required factorisation property.


The second approach  will be  based on direct integration over 
the $H=U(1)$ gauge 
fields starting with   \eqref{gwzwact}.
 The resulting quantum determinant  computed 
following  \ci{st} will   produce  a   local counterterm
which will contribute to the S-matrix.  
Note that neither of these two approaches  can be 
used to compute the S-matrix  in  the case of a 
non-abelian gauge group $H$:   they 
involve fixing a gauge on $g$, which cannot be done in a 
non-singular way  when expanding the action near $g=1$ 
 for   non-abelian $H$  (see \ci{ht}).

The third approach  \ci{ht}, which can  be  used for a non-abelian $H$, 
 will be  based on
the gauge choice  $A_+=0$. It  was already  described in
general  in section \ref{rev}. 

\subsubsection{Approach based on quantum effective action of 
gauged WZW theory\la{v1}}


The motivation  for the approach 
in  this subsection will be partly heuristic. 
We shall  start with   the local part of 
the quantum effective action for the  gauged WZW theory constructed in 
\ci{t1} and add to it  the same  potential  as in \rf{gwzwact}.
Even though the local part of the effective   action is 
formally  not  gauge invariant, we
will insist  that  it should  describe the same massive
degrees of freedom  which were present at 
 the classical level, i.e. we will still 
 parametrise $g$ as in \eqref{param} and then 
integrate out $A_\pm$.
We will then compute the resulting one-loop 
S-matrix.\foot{Though we start with
 the effective action, this effective action is 
 by construction an action for the  current variables rather than $g$. 
 Also, we omit the non-local contributions. For that reason 
we are still to include quantum loop contributions coming from the classical
 part of the effective Lagrangian.}



In the case when  $H$ is abelian the local part of the 
quantum effective action of the (axially) gauged WZW theory \ci{t1}
supplemented with  the  potential is (cf. \rf{gwzwact}; here $\nu= \ha$)\foot{
Here $\dcn$ is the dual Coxeter number  of $G$, i.e. the  value of the
second Casimir operator in the adjoint representation. Note that in \ci{t1} 
$c_G$ is defined as $2\dcn$. Also, 
 $\del$ and $\bar{\del}$ in
\ci{t1} are  respectively  $\fr{\dpl}{2}$ and $\fr{\dm}{2}$ here and
similarly for the gauge field components.} 
\bea
\mc{S}_{\rm eff} &\!\!\!  = & \!\!\!   \no
    - \  \fr{k+\dcn}{4\pi}   \trm{Tr} \Big[ \ \fr{1}{2} \,\int d^2x \;
      \ \inv{g}\dpl g\ \inv{g}\dm g\
    - \fr{1}{3}\,\int d^3x \;
      \ \epsilon^{mnl}
      \ \inv{g} \del_m g\ \inv{g}\del_n g \ \inv{g}\del_l g
\\    && \;\;\;\;\;\no
    + \,\int d^2x \; 
      \ \big[ A_+\dm g\inv{g} - A_-\inv{g}\dpl g 
             - \inv{g} A_+g A_- 
             - \big(1-\fr{2\dcn}{k+\dcn}\big) A_+A_- \big]
\\    && \;\;\;\;\; 
    + \ \gok\int d^2x \; 
      \ (\inv{g} T g T - T^2) \Big]\,.\la{gwzwactef}
\eea
To keep the mass of the elementary excitation as $m$ we assumed that the 
coefficient of the potential term is 
also shifted  from $k$ to $k+\dcn$. While we conjecture that the 
above action is correct 
to one-loop order, there may be further potential (or ``mixed'') 
 corrections depending on $m$ 
at higher orders.

In the  case of our present interest   $G=SU(2), \ H=U(1)$, i.e.  we have $\dcn=2$. 
Using the  
parametrisation of $g$ in \eqref{param} and solving for 
$A_\pm$  we  then arrive at the following effective 
Lagrangian 
\be\la{csgqu}
\mc{L}_{\rm eff} = \fr{k+2}{4\pi}\bigg[
                     \dpl \phi \dm \phi 
                   + \fr{\tan^2\phi \  \dpl \chi \dm \chi}
                        {1-\fr{2}{k}\tan^2\phi}
                   + \fr{m^2}{2}
                     (\cos2\phi-1)\bigg]\,, 
\ee
 $m$ is  then the  mass of the
 elementary excitations near the $\phi=0$ vacuum.
Rescaling $\phi$  to put  the kinetic part of the quadratic
Lagrangian in  canonical form  and expanding in 
  $\fr{1}{k}\ll 1$  gives
\be\la{csgquex}\begin{split}
\Lag_{\rm eff} = & \fr{1}{2}\Big(\dpl\phi\dm \phi
       + \phi^2\dpl \chi \dm \chi
       - m^2 \phi^2 \Big)
\\     & + \fr{2\pi}{k}\Big(\fr{1}{3}\phi^4
                             \dpl\chi \dm\chi
         + \fr{m^2}{6}\phi^4\Big)
\\     & + \fr{4\pi^2}{k^2}
           \Big(\fr{1}{6\pi}\phi^4\dpl\chi\dm \chi
               - \fr{m^2}{6\pi}\phi^4
               + \fr{17}{90}\phi^6\dpl\chi\dm\chi
               - \fr{m^2}{45}\phi^6\Big) 
               + \ord{k^{-3}}\,.
\end{split}\ee
Compared to the similar expansion of the original CsG  Lagrangian   \eqref{csgpc}  
we have additional $\fr{1}{k^2}$   terms   that may be
interpreted as 
quantum ``counterterms'' (cf. \rf{dvmcounter}) 
 required for  maintaining the 
integrability at the quantum level. 

Since $\Lag_{\rm eff}$ has  the  $ \chi \ra \chi+\al$ symmetry, 
introducing the ``cartesian'' coordinates (analogous to
$\psi_a$  in \eqref{lagcsg},\eqref{lagcsgex}) 
\be
Y_1=\phi \cos\chi\,, \hs{30pt} Y_2=\phi \sin\chi\,,
\ee
we may write the   Lagrangian in a manifestly $SO(2)$ invariant
form.
Computing   the perturbative one-loop  S-matrix  we get (cf.
\rf{smatso2},\rf{sesd})  
\be\begin{split}
\bra{Y_c(-p_3)Y_d(-p_4)}\mbb{S}
                 \ket{Y_a(p_1)Y_b(p_2)}&\no
\\=(2\pi)^24\ep_{p_1}\ep_{p_2}\de(p_1+p_3)
&\de(p_2+p_4) \ \big[\hS_1(\q)\ \de_{ab}\de_{cd} 
+ \hS_2(\q)\ \de_{ac}\de_{bd}+\hS_3(\q)\  
\de_{ad}\de_{bc}\big]\,,
\end{split}\ee
\be\la{smatcsgqu}\begin{split}
\hS_1(\q) = & - \fr{2\pi i}{k} \coth \q  
                 + \fr{4\pi^2}{k^2} \coth \q \cosech \q 
                 + \ord{k^{-3}}\,,
\\ \hS_2(\q) = & 1 + \fr{2\pi i}{k} \cosech \q 
                   - \fr{4\pi^2}{k^2}
                    (\fr{1}{2}+\cosech^2\q)
                   + \ord{k^{-3}}\,,
\\ \hS_3(\q) = & \fr{2\pi i}{k} \coth \q 
                   - \fr{4\pi^2}{k^2} \coth \q \cosech \q 
                   + \ord{k^{-3}}\,.
\end{split}\ee
We conclude that the resulting one-loop S-matrix  has the
factorisation property 
  as the reflection coefficient \eqref{r} vanishes.
Also,   $\hS= \hS_2(\q)+\hS_3(\q)$  (defined as in \rf{tdsof},\rf{spm}) 
agrees precisely  with the expansion \eqref{smatsolex} of the exact 
 S-matrix  \rf{smatsol} of \ci{dh}.

\subsubsection{Approach based on  direct integrating out
of
$A_+,A_-$\la{v2}}

Let us now show that  we can get the same one-loop 
 S-matrix by starting with the  $SU(2)/U(1)$  action 
 \eqref{gwzwact}  and  directly 
integrating out $A_\pm$ in the path integral, 
 taking into account 
the corresponding   determinant contribution.

Fixing  the same  gauge on  $g$ as
in \eqref{param}  and setting  
\be
A_\pm = a_\pm \ta_3\,,
\ee
the resulting action  becomes 
\bea
\mc{S} & = & \fr{k}{4\pi}\,\int d^2 x \; 
        \Big[ \dpl \phi \dm \phi 
       + \sin^2 \phi \; \dpl \chi \dm \chi \no
    \\\la{act1}  && \hs{23pt} -\  a_+ \sin^2\phi \; \dm \chi 
               - a_- \sin^2\phi \; \dpl \chi 
               - a_-a_+\cos^2\phi 
               + \fr{m^2}{2}(\cos2\phi-1)\,\Big] \ .
\eea
 If we simply solve for the gauge field components 
 $a_\pm$   we will then 
arrive at the Lagrangian  in \eqref{csgpc}. However, integrating out $a_\pm$ in the path integral 
requires careful definition of the measure and may result in a non-trivial quantum determinant.
In addition to the well-known dilaton term on a  curved 2-d
background \ci{busher,duad}
there is also a  local 2-derivative   contribution  \ci{duad} discussed
in detail in the appendix of
\ci{st}.
In general, starting  with a path integral of the form
\be\la{paintst}
\mc{Z} = \int [da] 
         \exp \Big[\fr{i}{2}\int d^2 x\; 
                    \M(\phi)\   a_+ a_-\Big]\,,
\ee
where $a_\pm$ is a 2-d vector field  and assuming a natural
definition 
of the resulting determinant (equivalent to setting $a_+ = \del_+ u, \ a_-= \del_- v$ and integrating over the scalar fields $u,v$)  one finds the following 
local contribution to the effective action 
\be\la{result}
-\fr{1}{8\pi}\int d^2x\; \dpl \ln \M              \    \dm 
\ln \M \,.
\ee
Noting that in the present case $\M= \cos^2 \phi$
we thus get an  extra ``counterterm''  that should be
added to the ``naive'' 
action \eqref{csgpc} 
\be\la{qct1}\begin{split}
& -\fr{1}{2\pi}\int d^2x \; 
  \dpl \ln \cos  \phi  \ 
  \dm  \ln \cos  \phi \,.
\end{split}\ee
The result is  the following ``corrected''  Lagrangian  (cf.
\eqref{csgqu}) 
\be\la{act3}
\mc{L}_{\rm corr}  = \fr{k}{4\pi} \bigg[ \dpl \phi \dm \phi
       + \tan^2 \phi\  \dpl \chi \dm \chi
       + \fr{m^2}{2}
         (\cos 2\phi-1)- \fr{2}{k}
           \tan^2\phi\ \dpl\phi\dm \phi\bigg]\,, 
\ee
where the   $k^{-1}$ term is the  one-loop  determinant  contribution.

To compute the S-matrix we again rescale $\phi$ by
$\sqrt{\fr{2\pi}{k}}$ 
and expand the Lagrangian as  (cf. \eqref{csgquex})
\be\la{csgquex1}\begin{split}
\Lag_{\rm corr} = & \fr{1}{2}\big( \dpl\phi\dm \phi
       + \phi^2 \dpl \chi \dm \chi
       - {m^2}\phi^2\Big)
\\   & + \fr{2\pi}{k}\Big(\fr{1}{3}\phi^4\dpl \chi \dm\chi
       + \fr{m^2}{6}\phi^4\Big)
\\   & + \fr{4\pi^2}{k^2}
         \Big(-\fr{1}{2\pi}\phi^2\dpl\phi\dm\phi
               +\fr{17}{90}\phi^6\dpl\chi\dm\chi
               -\fr{m^2}{45}\phi^6\Big)
       + \ord{k^{-3}}\,.
\end{split}\ee
One can   immediately see that this Lagrangian  gives 
the same one-loop two-particle S-matrix as the Lagrangian in 
$\eqref{csgquex}$  as they are related by a field
redefinition
\be\la{frd}
\phi \ra \phi -\fr{2\pi}{3k^2}\phi^3
                    +\ord{k^{-3}}\,, 
\hs{20pt} \ \ \ \ \   \chi\ra \chi\,.
\ee
Thus the one-loop S-matrix computed using  \eqref{act3}
again 
agrees with \rf{smatsolex}, i.e. with 
the 
 exact 
S-matrix  of \ci{dh}.

As we will  show in the appendix, 
the field redefinition \eqref{frd} can be extended to higher orders in 
the field $\phi$  relating the two closed-form
Lagrangians \eqref{csgqu} and \eqref{act3}  to one-loop order, i.e. 
up to $\fr{1}{k^2}$ terms. 

\subsubsection{Approach based on $A_+=0$  gauge 
	    \la{v3}}

Next, let us consider 
 the computation of the one-loop S-matrix using the $A_+=0$ gauge
approach described  already 
in section \ref{rev}
 for the case of  $G/H=SU(2)/U(1)$.  
In terms of the generators $\tau_A$ in  \eqref{gen}
we 
define  the normalised (as in \eqref{noo}) 
$G=SU(2)$ generators (cf. \eqref{gen})
\be\la{gensu22}
\TT_A= 2\sqrt{\fr{2\pi}{k}} \tau_A\,, \ \ \ \ \ \ \ \ \   K=
\fr{4 \pi}{k}  \ . 
\ee
We then find that   the component fields  $X_a$  ($a=1,2$) 
defined as in
\eqref{cx}  have canonical  kinetic
terms in the  action  following from  \eqref{gwzwex3}  ($T=\tau_3$)  
\bea
\mc{S}[X] 
      & \!\!\!=\!\!\! & \int d^2x \; \Big[\ 
          \fr{1}{2}\Big( \dpl X_a \dm X_a   
        - {m^2} X_a X_a \Big) \no
\\&&  \   +\   \fr{2\pi}{3k}\Big(X_a X_a
        \dpl X_b \dm X_b 
        -  X_a \dpl X_a X_b \dm X_b  
 +\ \fr{m^2}{2}X_a X_a X_b X_b\Big)
              + \ord{k^{-2}}\Big] \,.\la{gwzwex3csg}
\eea
This action agrees with the expansion of the 
complex sine-Gordon action \eqref{csgpc} to quartic order, up to a 
field redefinition. Therefore, its  contribution to 
 the one-loop S-matrix will be the same  as the direct 
CsG  model  result \eqref{smatcsg}  with the
coupling $\zg^2$ replaced by $\fr{2\pi}{k}$. 

In addition to the contribution of the action 
 we should include the 
contribution of the functional determinant \eqref{op4}  in
\eqref{nabcont}.
Using the generators in \eqref{gensu22} to define the
structure constants,  we have the following relations 
for $V_{abij}=f_{aic} f_{bjc}$  in \eqref{op4}
\be\la{su2ident}\begin{split}
V_{(ab)ij}V_{(cd)ji} = 
          \fr{64\pi^2}{k^2}\de_{ab}\de_{cd}\ ,
\  \ \ \ \ \ \ \ \ \ \ \ \ \ \ V_{[ab]ij}V_{[cd]ji} = 0\,.
\end{split}\ee
Substituting them into \eqref{nabcont} gives the following determinant 
contributions to 
the one-loop S-matrix  coming 
from the $s,t,u$-channel diagrams in figure
\ref{fig2}
\be\la{smatcorr}
  \Delta S_1=  -\fr{2\pi i}{k^2}(\cosech\q+\coth\q)\,; \ \  \ 
  \Delta S_2=  0 \, ; \ \ \  
   \Delta S_3=
 -\fr{2\pi i}{k^2}(\cosech\q-\coth\q)\, .
  \ee
Summing these up  with 
 the direct  CsG  result
\eqref{smatcsg} we are led 
again to the same reflectionless 
one-loop S-matrix matching 
 the expansion \eqref{smatsolex}
 of the exact  S-matrix of \ci{dh}.


\subsection{Comments\la{comments}}

The three methods discussed in this section  all gave  
the same one-loop correction to the perturbative S-matrix \eqref{smatcsg}
of the complex sine-Gordon theory defined by the Lagrangian
\eqref{lagcsg}.  
This correction originated from the  definition of 
 the quantum 
complex sine-Gordon theory in terms of the $SU(2)/U(1)$
gauged WZW model with a potential.

 The first method was based on  starting with 
  the local part of the quantum effective action of
the gauged WZW theory,  while the
second and third methods were based 
on taking into account the  quantum corrections
 to the process of gauge fixing and 
integrating out the gauge fields in the path integral. 


As was mentioned in section \ref{v2} and explicitly shown in the
appendix,  the closed-form Lagrangians of the first
two methods, \eqref{csgqu} and \eqref{act3}, are related by
a field redefinition 
if considered to the leading 
(one-loop) order in $\fr{1}{k}$. 
The resulting one-loop S-matrix  is consistent with factorisation 
and  agrees with the exact solitonic S-matrix of \ci{dh}.

It remains to be seen  
if this agreement persists  beyond the one-loop order. 
The quantum effective action of gauged WZW theory employed
in section \ref{v1} is by construction  consistent with 
quantum  conformal
symmetry. However, the presence of the potential 
leads to an extra renormalization  and its effect  on the full  theory is
to be understood better.  
To this end one may  study  corrections to the
mass  $m$  of the elementary excitations at higher  loops. 
For consistency with the results of
\ci{dvm2,dh}
these should 
match the expansion of the quantum soliton mass \eqref{solmassk}
in the special case of  $Q=1$.



\renewcommand{\theequation}{4.\arabic{equation}}
\setcounter{equation}{0}
\section{Perturbative one-loop S-matrix 
of  $SO(N+1)/SO(N)$ \\
gauged WZW theory 
with integrable potential \la{sec4}}

The most
important advantage of the third approach  discussed in the previous section
and in section 2, i.e. fixing the $A_+=0$  gauge
and integrating out $A_-$, is that it can be applied to the 
case of $G/H$ gauged WZW theory with a  non-abelian gauge group $H$.
Below we  shall use this method   to compute the one-loop
perturbative S-matrix for the $SO(N+1)/SO(N)$  theory.

\subsection{Basic definitions \la{group}}


Our starting point  will be the  path integral 
for the  $G/H$  gauged WZW theory with integrable
potential 
  \eqref{gwzwact}.
In general,  to define that  
 $G/H$ theory, one may 
 consider $G$ embedded into a larger group $ F$ (see  \ci{bakps,gt,mirp}).
The matrix $T$ then lives in the orthogonal complement $\mf{p}$ of 
$\mf{g}$ in the algebra $\mf{f}$ of  $F$.
 $T$ is chosen to be an element of
the  maximal abelian subalgebra $\mf{a}$ of $\mf{p}$
which we shall assume to be one-dimensional. $H$ 
is then defined as the maximal subgroup of $G$ satisfying 
$\com{\mf{h}}{T} = 0$.
Consequently,  the potential in \eqref{gwzwact}  preserves the local 
$H$
symmetry of the gauged WZW theory.
In this case the elementary  excitations of the
action   \eqref{gwzwact}  near the  $g=1$ vacuum point are all  massive.

In the case of $G/H=SO(N+1)/SO(N)$, we  choose  $F=SO(N+2)$.
$g$ is taken to be  a matrix  in the fundamental
representation of $G=SO(N+1)$  so that $\ind=1$ in \eqref{gwzwact}.
  We choose the
following standard basis for
the fundamental representation of $F=SO(N+2)$ 
represented by 
$(N+2) \x (N+2)$ real antisymmetric matrices
(see section
2 and 
 also \ci{ht})
\be\la{qr}
\ba{c}
\TT_{\hat{A}} = \{\TTm_{\al\bet}: \al<\bet\}\,, \hs{20pt} 
              (  \TTm_{\al\bet})_{xy} = \fr{K}{\sqrt{2}}
              (  \de_{\al x}\de_{\bet y}
               - \de_{\bet x}\de_{\al y})\,,
\\ \al,\,\bet,\,x\,,y = -1,\,0,\ldots,\,N\,.
\ea\ee
The index $\hat{A}=1,\ldots,\,\fr{(N+2)(N+1)}{2}$ labels the
generators of $\mf{f}$. 
$\mf{g}=\so(N+1)$ is then  the subalgebra generated by
elements of $\mf{f}=\so(N+2)$ that are non-zero in the bottom-right
$(N+1) \x (N+1)$ corner
\be\la{rqr}\ba{c}
\TT_{A} = \{\TTm_{\al\bet}: \al<\bet\}\,, \hs{20pt} 
        (  \TTm_{\al\bet})_{xy} = \fr{K}{\sqrt{2}}
        (  \de_{\al x}\de_{\bet y}
         - \de_{\bet x}\de_{\al y})\,,
\\ \al,\,\bet = 0,\ldots,\,N\ , \hs{20pt}
   x,\,y = -1,\,0,\ldots,\,N\,.
\ea\ee
The matrix 
$T$ in the potential  may be chosen as 
\be\la{qrqr}
T = \fr{\sqrt{2}}{K}\TT_{-1\ts,\ts 0}\,.
\ee
The normalisation is fixed so that the  mass of the
elementary excitations in \eqref{gwzwact} is given by $m$.
This choice of $T$ then specifies the generators of the
algebra
$\mf{h}=\so(N)$ to be those elements of $\mf{g}=\so(N+1)$
that are non-zero in the bottom-right $N \x N$ corner
\be\la{hgen}\ba{c}
\TT_{i} = \{ \TTm_{ab}: a<b\}\,, \hs{20pt} 
        (  \TTm_{ab})_{xy} = \fr{K}{\sqrt{2}}
        (  \de_{a x}\de_{b y}
         - \de_{b x}\de_{a y})\,,
\\ a,\,b = 1,\ldots,\,N\ , \hs{20pt}
   x,\,y = -1,\,0,\ldots,\,N\,.
\ea\ee
Finally,  the basis for the ``physical'' 
 coset part  $\mf{m}$ is given by
\be\la{mgen}\ba{c}
\TT_{a} = \{ \TTm_{0a}\}\,,\hs{20pt} 
        (  \TTm_{0a})_{xy} = \fr{K}{\sqrt{2}}
        (  \de_{0 x}\de_{a y} - 
           \de_{a x}\de_{0 y})\,,
\\a = 1,\ldots,\,N\ , \hs{20pt}
  x,\,y = -1,\,0,\ldots,\,N\,.
\ea\ee
It is fairly easy to see that all these bases satisfy the 
requirements  listed  at the beginning of
section \ref{rev}.

As discussed in section \ref{rev}, the one-loop S-matrix computed in 
 the gauge $A_+=0$ receives contributions directly from the 
 vertices in the action  
\eqref{gwzwex3}  but   also   from the  
determinant \eqref{op}. Let us  consider these two types of 
contributions in turn.

\subsection{Lagrangian contribution}

We start with the gauge-fixed action 
 \rf{gwzwex3}   and set $X =  X_a \TT_a$   
 as in  \eqref{cx}. Then 
$X_a$ 
will have canonical  kinetic and mass
terms provided   we choose the normalization constant in \rf{qr}--\rf{mgen} 
as
\be\la{constants}
   K = 2\sqrt{\fr{2\pi}{k}}\,.
\ee
Then \eqref{gwzwex3}  gives   the following 
 quartic action for $X_a$  
\bea
\mc{S}[X] 
     & \!\!\!=\!\!\! & \int d^2x \; \Big[\;
         \fr{1}{2}\Big( \dpl X_a \dm X_a 
       - {m^2} X_a X_a\Big) \no
\\ && \hs{5pt} +\  \fr{\pi}{3k}\Big( X_a X_a 
                                   \dpl X_b \dm X_b 
       -  X_a \dpl X_a X_b \dm X_b  +   \fr{m^2}{2}X_a X_a X_b X_b\Big)
                + \ord{k^{-2}}\, \Big],\la{gwzwex4}
\eea
where the  indices $a,\,b,\ldots=1,\ldots,\,N$  
are contracted with $\de_{ab}$.
Note that in the $N=2$   case, i.e. when  $G/H=SO(3)/SO(2)$,
this 
should lead to the same  action as in  \eqref{gwzwex3csg} 
corresponding to the  case of 
 $G/H=SU(2)/U(1)$.
 Indeed, the two  actions are related by the rescaling  $k \to 2k$.\foot{This 
  rescaling may be attributed to the fact  that $SU(2)$ is 
 double cover of
$SO(3)$. Note also that the dual
Coxeter number of $SU(2)$, i.e.   $\dcn=2$, is  twice that
of $SO(3)$, i.e.  $\dcn=1$.}


The  $SO(N)$ S-matrix has the same  structure \eqref{smatso2},\rf{sesd} 
  as in  the
$SO(2)$ case, now  with the
indices $a,\,b =1,\ldots,\,N$. 
The contributions  to the one-loop 
S-matrix 
coming directly from 
 the Lagrangian \eqref{gwzwex4}  are then given by  \ci{dvm1}
\be\la{smatnab}\begin{split}
{S}_1(\q,N) = &-\fr{i\pi}{k} \coth \q 
                   + \fr{i\pi}{2k^2}(\cosech \q + \coth \q)
                   + \fr{\pi^2}{k^2}\coth \q \cosech \q 
\\ & \hs{150pt}  
  + \fr{i\pi}{2k^2}(N-2)(i\pi-\q)\coth^2\q
                   +\ord{k^{-3}}
\,,
\\ {S}_2(\q,N) = & 1 
                  + \fr{i\pi}{k} \cosech \q 
                  - \fr{\pi^2}{k^2}(\fr{1}{2}+\cosech^2\q) 
                  + \fr{i\pi}{2k^2} (N-2) \cosech \q 
                  +\ord{k^{-3}}
\,,
\\ {S}_3(\q,N) = & \fr{i\pi}{k} \coth \q 
                  + \fr{i\pi}{2k^2}(\cosech \q - \coth \q)
                  - \fr{\pi^2}{k^2} \coth \q \cosech \q
\\ & \hs{150pt}
   + \fr{i\pi}{2k^2}(N-2)\q\coth^2\q 
                  +\ord{k^{-3}}
\,.
\end{split}\ee

\subsection{Determinant contribution}

Next, let us  compute   the contribution to the one-loop S-matrix 
\eqref{op4},\eqref{nabcont}  coming from  the determinant
\eqref{op}, resulting from integrating out $A_-$ and solving
for $\xi$ in \rf{conex}. 
 In section \ref{revfd} this
contribution \eqref{nabcont} was computed for generic 
 structure constants (or arbitrary 
$V_{abij}=f_{aic}f_{bjc}$).
For the present $SO(N+1)/SO(N)$ case 
with \rf{qr}-\rf{mgen},\eqref{constants} 
we have the following  relations 
\be\la{sonident}\begin{split}
V_{(ab)ij}V_{(cd)ji} = & 
          \fr{16\pi^2}{k^2}\Big[\de_{ab}\de_{cd}
         +\fr{1}{2}(N-2)(\de_{ad}\de_{bc}
                        +\de_{ac}\de_{bd})\Big]\,,
\\V_{[ab]ij}V_{[cd]ji} = &
          \fr{16\pi^2}{k^2}\Big[
          \fr{1}{2}(N-2)(\de_{ad}\de_{bc}
                        -\de_{ac}\de_{bd})\Big]\,.
\end{split}\ee
Using these  in \eqref{nabcont} we conclude  that the 
contribution of the determinant \eqref{op} to the three 
functions ${S}_1, \, {S}_2$ and $ {S}_3$ in \rf{sesd} are, respectively, 
\be\la{contdetnab}\begin{split}
\Delta S_1(\q,N)= -\fr{i\pi}{2k^2} & (\cosech\q+\coth \q) 
                 + (N-2)\fr{i\pi}{k^2}\coth \q\,,
\\\Delta S_2(\q,N)= -\fr{i\pi}{k^2} & (N-2)\cosech\q\,,
\\\Delta S_3(\q,N)= -\fr{i\pi}{2k^2} & (\cosech\q-\coth\q)
                   - (N-2)\fr{i\pi}{k^2}\coth\q\,.
\end{split}\ee

\subsection{One-loop  S-matrix and the Yang-Baxter equation}

Summing up \eqref{smatnab} and \eqref{contdetnab} we get  the
following expression for the one-loop S-matrix of the 
 $SO(N+1)/SO(N)$ theory 
\be \la{too}
\hS_i=  S_i + \Delta S_i\ , \ \ \ \ \ \ \ \ 
\hS_{ab}^{cd} =  \hS_1\ \delta_{ab} \delta^{cd}\, + 
\hS_2\ \delta_{a}^c \delta^{d}_b \,
+  \hS_3\ \delta_{a}^d \delta^{c}_b \ , \ee
where 
\be\begin{split}
\hS_1(\q,N) = & 
                     - \fr{i\pi}{k} \coth \q 
                     + \fr{\pi^2}{k^2}\coth \q \cosech \q 
\\       & \hs{50pt} 
+ \fr{i\pi}{2k^2}(N-2)
           \big[(i\pi-\q)\coth^2\q+2\coth\q \big] 
                     +\ord{k^{-3}}\,,
\\
\hS_2(\q,N) = & 
                  1 + \fr{i\pi}{k} \cosech \q 
                    - \fr{\pi^2}{k^2}
                     (\fr{1}{2}+\cosech^2\q) 
                    -\fr{i\pi}{2k^2} (N-2) \cosech \q 
                    +\ord{k^{-3}}\,, \la{sm}
\\
\hS_3(\q,N) = & 
                       \fr{i\pi}{k} \coth \q 
                     - \fr{\pi^2}{k^2} \coth \q \cosech \q
\\       & \hs{50pt} 
+ \fr{i\pi}{2k^2}(N-2)
           \big[\q\coth^2\q-2\coth\q \big]
                     + \ord{k^{-3}}\,.
\end{split}\ee
For   $N=2$ this  expression  agrees with
the expansion  of the exact S-matrix \ci{dh} of the complex sine-Gordon theory
in \rf{smatsol},\rf{smatsolex}: with  $k$  rescaled by $\fr{1}{2}$
 \eqref{sm} reduces to \eqref{smatcsgqu}.

\

Let us  now study whether  this S-matrix satisfies  the
Yang-Baxter equation (YBE). 
Let us  define the following  tensor function 
\be\la{yb}
Y_{abc}^{def}(\q_{12},\,\q_{23}) = 
\hS_{ab}^{gh}(\q_{12})\ 
\hS_{gc}^{du}(\q_{13})\ 
\hS_{hu}^{ef}(\q_{23}) -
\hS_{bc}^{hu}(\q_{23})\ 
\hS_{au}^{gf}(\q_{13})\ 
\hS_{gh}^{de}(\q_{12})\ ,
\ee
where $a,\,b,\ldots=1, \ldots,\, N$ and 
 $\q_{ij}=\q_{i}-\q_{j}$, with  $\q_i$  being 
the rapidities of the three particles being scattered.
 The condition of factorisation for an S-matrix of 
a standard Lorentz-invariant integrable theory is that 
it should  satisfy the 
quantum YBE, i.e. that 
 the tensor function 
$Y_{abc}^{def}$ should vanish.

If we  define the leading non-trivial 
term in  the large $k$ expansion  of  $Y_{abc}^{def}$ 
 (the order $\fr{1}{k}$ term trivially vanishes)
\be\la{cyb}
y_{abc}^{def}=Y_{abc}^{def}\Big|_{\ord{k^{-2}}}\,, 
\ee
then its vanishing, i.e. $y_{abc}^{def}=0,$
is equivalent to the condition that the
 tree-level S-matrix satisfies the 
classical Yang-Baxter equation.

When  $a,\,b,\ldots$ are $SO(N)$ vector indices, 
the tensor $Y_{abc}^{def}$ can be parametrised by a set of 15 
functions $A_I$  as follows
\be\la{qwe}\begin{split}
Y_{abc}^{def} (\q_{12},\,\q_{23})
= &
A_1\ \de_a^d\de_b^e\de_c^f + 
A_2\ \de_a^d\de_b^f\de_c^e +
A_3 \ \de_a^f\de_b^e\de_c^d +
A_4\ \de_a^e\de_b^d\de_c^f \\&\quad +
A_5\ \de_{ab}\de^{de}\de_c^f +
A_6\ \de_{ac}\de^{df}\de_b^e +
A_7\ \de_{bc}\de^{ef}\de_a^d \\&\quad +
A_8\ \de_{ab}\de^{ef}\de_c^d +
A_9\ \de_{bc}\de^{de}\de_a^f +
A_{10}\ \de_{ac}\de^{de}\de_b^f +
A_{11}\ \de_{ab}\de^{df}\de_c^e \\&\quad +
A_{12}\ \de_{bc}\de^{df}\de_a^e +
A_{13}\ \de_{ac}\de^{ef}\de_b^d +
A_{14}\ \de_a^f\de_b^d\de_c^e +
A_{15}\ \de_a^e\de_b^f\de_c^d \,, 
\end{split}\ee
where 
\be 
A_I = \fr{\pi^2}{k^2} A^\nb{2}_I
 + \fr{\pi^3}{k^3} A^\nb{2}_I+\ord{k^{-4}}\,. \la{rer}
\ee
 Here   $A^\nb{2}_I$  are the  tree-level   and  
  $A^\nb{3}_I$ are the one-loop  coefficient functions
  which, in general, are functions of $\q_{12}$ and $\q_{23}$.
  

Remarkably,  for the S-matrix \rf{too}   parametrised
by the functions
in \eqref{sm}, the tree-level coefficients 
$A^\nb{2}_I$ turn out to be   simple constants
\be\la{klp}\begin{split}
  A^\nb{2}_1 = A^\nb{2}_2 & 
= A^\nb{2}_3 = A^\nb{2}_4 
= A^\nb{2}_5 = A^\nb{2}_6 
= A^\nb{2}_7 = 0\,,
\\ & A^\nb{2}_8 = A^\nb{2}_{10} 
   = A^\nb{2}_{12} = A^\nb{2}_{14} = 1\,,
\  \ \ \ \ \ \ \  A^\nb{2}_9 = A^\nb{2}_{11} = 
     A^\nb{2}_{13} = A^\nb{2}_{15} = -1\,.
\end{split}\ee
The vanishing of the first seven functions extends  also to the 
one-loop order, i.e. 
\be\la{hoh}
  A^\nb{3}_1 = A^\nb{3}_2 = A^\nb{3}_3 
= A^\nb{3}_4 = A^\nb{3}_5 = A^\nb{3}_6
= A^\nb{3}_7=0\,.
\ee
However, this does not apply to the remaining 8 functions. 
 They can, however,  be written in
a compact form using the following three functions $f_1,
f_2, f_3$  of the rapidities 
\be\la{koi}\begin{split}
f_1(\q_{12},\,\q_{13},\,\q_{23}) \equiv  & 
            \cosech\q_{12}\cosech\q_{13}\cosech\q_{23}\,,
\\f_2(\q_{12},\,\q_{13},\,\q_{23}) = &
              \fr{i}{2}f_1(\q_{12},\,\q_{13},\,\q_{23})
\\&\times \Big[\cosh(\q_{12}+\q_{13}) 
      + \cosh(\q_{12}+\q_{23}) 
      + \cosh(\q_{13}+\q_{23})
\\&\hs{20pt} - \cosh(\q_{12}-\q_{13}) 
             - \cosh(\q_{12}-\q_{23}) 
             - \cosh(\q_{13}-\q_{23})\Big]\,,
\\f_3(\q_{12},\,\q_{13},\,\q_{23}) = & 
     -\fr{1}{4\pi}f_1(\q_{12},\,\q_{13},\,\q_{23})
\\&   \hs{40pt}\times \Big[\q_{12}\cosh2\q_{12}
                   +\q_{13}\cosh2\q_{13}
                   +\q_{23}\cosh2\q_{23}\Big]\,.
\end{split}\ee
Namely, 
\bea
A^\nb{3}_8 = - A^\nb{3}_9  \no
          &\!\!\! = \!\!\!& f_2(\q_{12},\,\q_{13},\,\q_{23})
\\&& \no -(N-2)\big(\fr{2}{\pi}
     - \fr{i}{4}f_1(\q_{12},\,\q_{13},\,\q_{23})
-f_3(\q_{12},\,i\pi-\q_{13},\,\q_{23})\big)\,,
\\A^\nb{3}_{10} = - A^\nb{3}_{11} \no
               & \!\!\!= \!\!\!&
f_2(\q_{12},\,\q_{13},\,\q_{23})-(N-2)\big(\fr{2}{\pi}
+f_3(-i\pi+\q_{12},\,i\pi-\q_{13},\,\q_{23})\big)\,,
\\A^\nb{3}_{12} = - A^\nb{3}_{13} \no  
               & \!\!\!= \!\!\!& f_2(\q_{12},\,\q_{13},\,\q_{23})
-(N-2)\big(\fr{2}{\pi}
+f_3(\q_{12},\,i\pi-\q_{13},\,-i\pi+\q_{23})\big)\,,
\\A^\nb{3}_{14} = - A^\nb{3}_{15} 
               & \!\!\!= \!\!\!& f_2(\q_{12},\,\q_{13},\,\q_{23})
-(N-2)\big( \fr{2}{\pi}
+f_3(\q_{12},\,-\q_{13},\,\q_{23})\big)\,. \la{wew}
\eea
Below we shall    discuss the properties of 
these coefficients in some special cases. 

First, in the abelian  $N=2$  case we 
note that $\de_{ab}\de_{cd}\de_{ef}$ and its permutations
are not independent. Therefore,  the functions $A_I$
  end up combining in such a way that
\be
Y_{abc}^{def}(\q_{12},\,\q_{23})=0+\ord{k^{-4}}\,.
\ee
As a result, in this case 
 the Yang-Baxter equation for the perturbative S-matrix 
is satisfied to the
one-loop  order we consider here. 
This is a manifestation of  the integrability 
of the  quantum  
complex sine-Gordon model 
discussed in section \ref{csgdvm}.

\subsubsection{Non-abelian  case $N\geq 3$: tree level }

 In the $N\geq 3$ case
the tree-level part 
of the l.h.s. of the 
Yang-Baxter equation
\eqref{cyb} 
can be written in the following  way
\be\la{ycom}\begin{split}
y_{abc}^{def} = & \fr{\pi^2}{k^2}\Big(
 \de_{ab}\de^{ef}\de_c^d -
 \de_{bc}\de^{de}\de_a^f +
 \de_{ac}\de^{de}\de_b^f -
 \de_{ab}\de^{df}\de_c^e  
\\ & \quad 
+
 \de_{bc}\de^{df}\de_a^e -
 \de_{ac}\de^{ef}\de_b^d +
 \de_a^f\de_b^d\de_c^e   -
 \de_a^e\de_b^f\de_c^d\Big) 
\\ =  &\  \fr{1}{16}\sqrt{\fr{\pi}{k}}\ 
       \Tr\big(\TTm_a^{\;d}
       \com{\TTm_b^{\;e}}{\TTm_c^{\;f}}\big)\,,
\end{split}
\ee
where $\TTm_a^{\;b}$ are the generators of  the algebra 
$\mf{h}=\so(N)$, defined in \eqref{hgen}, 
so that $y_{abc}^{def}$ is just proportional to the structure constants 
of $\mf{h}$.\foot{The constant
$K$ in \eqref{hgen} is defined in terms of $k$ in
\eqref{constants} and the indices $a,\,b,\ldots$ are
raised and lowered with $\de_{ab}$, $\de^{ab}$.} 

We conclude that  while in the non-abelian 
case the tree-level S-matrix does not satisfy the
standard  Yang-Baxter equation, 
the ``anomaly'' has a remarkably simple form: 
it is independent of the rapidities  and 
is proportional to  the {\it structure
constants} of the $\so(N)$ algebra.\foot{Note that  the same  expression for  $y_{abc}^{def}$
is found   if we  formally compute it using 
the following constant matrix $(S_0)_{ab}^{cd}$ 
instead of $\hat S_{ab}^{cd}$ in \eqref{sm}:
\be\no (S_0)_{ab}^{cd} = \de_a^c\de_b^d + t_{ab}^{cd} \ , \ \ \ \ \ \ 
t_{ab}^{cd} = \fr{i}{8}\Tr(\TTm_a^{\;c}\TTm_b^{\;d})
            = \fr{i\pi}{k}(\de_a^d\de_b^c
                         - \de_{ab}\de^{cd})\,.
\ee
}

The  violation of the classical YBE  for the $N \geq 3 $
S-matrix in  \eqref{sm}  could have been  expected as it has a
non-trivial ``trigonometric'' dependence 
on the rapidities. At the same time,  a well-known fact is 
that  
a tree-level S-matrix with a non-abelian symmetry 
satisfying the YBE  must have a  rational form \ci{zz}. 
It is  remarkable, however,   that  the  
 S-matrix in  \eqref{sm} 
   violates the classical YBE by only a
constant term 
 proportional to 
the structure
constants  of the  global symmetry algebra.
This   suggests that the satisfaction 
of the YBE may be restored  by some
modification or re-interpretation of the S-matrix 
(like a change of basis of states or a similarity transformation).

Indeed, the violation of the classical YBE appears to be
 in contradiction with 
the classical integrability of the theory \eqref{gwzwact}.
A possible  resolution  of this contradiction may 
be related to the fact  that the global $SO(N)$  symmetry 
of the $A_+=0$ gauge-fixed theory \eqref{gwzwex4}  and
thus 
of the  corresponding S-matrix is  actually unphysical:
this symmetry 
 is the  global part of the $H$ gauge  symmetry  that 
is not fixed by  the $A_+=0$ gauge.\foot{It is thus  a direct
analog of the usual global
$SU(N)$ symmetry  acting on on-shell  gluons in the  
formal perturbative QCD  S-matrix, with $X_a$ being the
analogs of gluons.
The
Yang-Baxter equation constraint  may  be weakened for such non gauge invariant 
excitations, e.g., it  might only be imposed  up  
to gauge transfomations on each of the legs (internal and 
external) in the usual three-particle scattering factorisation.}

Another indication that an apparent 
  violation of integrability is related to gauge fixing 
  is the following.  While the gauge-fixed action  \rf{gwzwex3} for $X_a$   formally 
admits a Lax  connection  \ci{ht}  and thus should still be
integrable, 
this Lax connection   contains non-local terms originating from 
the  gauge fixing procedure.
This may lead to a problem 
with  the standard derivation
of the YBE. It is natural to expect 
that this violation of the YBE   should be ``mild'', and this is indeed 
what we  found in 
\eqref{ycom}.

\


Another   idea  of how  to try   to  ``repair'' the violation of the YBE 
 is  based on the observation 
that in models such as generalised sine-Gordon models  based on 
 integrable deformations of gauged WZW theories  there are
hints of a hidden quantum group structure, with the 
symmetry group appearing  to be  broken to the
Cartan subgroup in the classical limit \cite{hmsol,beisctr}.
We may then  look for a constant 
tensor $s^{cd}_{ab}$ constructed out of $SO(2)$ invariants (i.e. breaking $SO(N)$
symmetry), such that
when we replace $\hS^{cd}_{ab}$  in \eqref{sm} by 
\be\la{jkjk}
\td S^{cd}_{ab}= \hS^{cd}_{ab} + s^{cd}_{ab}\,,
\ee
the classical Yang-Baxter equation will be  satisfied, 
$
y_{abc}^{def}=0.$
To demonstrate how this  idea  may work\foot{It is not 
clear if this procedure is unambiguous as it may depend 
on a choice of a particular  Cartan subgroup.}
let us  define 
  the indices $a_i$ (for $i$ odd and $1\leq i < N$, for  $N$
even)\foot{For odd $N$ we will need also to add $a_N$  taking the single value $N$.}
which take  values $i$ and $i+1$, 
i.e.   a  tensor carrying an index $a_i$ is  non-zero only when
this  index takes the value $i$ or $i+1$.
This choice of indices 
corresponds to 
selecting  the Cartan subgroup to be $2\x 2$ block diagonal 
in the usual matrix representation of $SO(N)$. The $SO(2)$
invariant tensors $\de_{ab},\,\ep_{ab}$ should carry a
pair of indices $a_i,\,b_i$ with the same value for $i$.
Defining the tensor 
\be
s_{ab}^{cd} = \fr{\pi}{k}
        \sum_{\substack{1\leq i<j\leq N\\i,j \trm{ odd}}}
\big(\ep_{a_ib_i}\de^{c_jd_j} 
    + \ep^{c_id_i}\de_{a_jb_j}
    - \ep_{a_i}^{\ d_i}\de_{b_j}^{c_j} 
    - \ep_{\ b_i}^{c_i}\de_{a_j}^{d_j}\big)\,, \la{hr}
\ee
one can check that $\td S^{cd}_{ab}= \hS^{cd}_{ab} + s^{cd}_{ab}$
indeed satisfies  the classical YBE.
The meaning of this observation still remains to be clarified. 

\subsubsection{The case of  $H=SO(4)$}

The case  of our prime interest is $G/H$=$SO(5)/SO(4)$ 
which is connected to $F/G$=$SO(6)/SO(5)$=$S^5$ and thus to the 
 bosonic string theory on $\mbb{R}_t \x S^5$
and  the  superstring theory on $AdS_5 \x S^5$. 
 As the vector
representation of $SO(4)$ is isomorphic to the bifundamental
representation 
of $SU(2) \x SU(2)$ we can rewrite the S-matrix \eqref{sm}  
in an  $SU(2) \x SU(2) $ covariant basis. 

Since here  the global symmetry  is a product group,  
the 
integrability implies that the S-matrix should
 factorise into a direct product of the 
two identical $SU(2)$ S-matrices, i.e. 
\be\la{fact}
  \hS_{\al\bet,\ald\betd}^{\g \de,\gad\ded}
= \hS_{\al\bet}^{\g\de}\ \hS_{\ald\betd}^{\gad\ded}\,,
\ee
where $\al,\,\bet,\ldots=1,\,2$ and $\ald,\,\betd,\ldots=1,2$
are the $SU(2)$ 
indices.\foot{This group factorisation is a non-trivial consequence of integrability: 
  for example,   the standard  $SO(4)/SO(3)$ 
geometric coset  sigma 
model with  a potential, studied in appendix D of
\ci{ht},  has  the same symmetry and field content but is
not integrable 
so that  its  S-matrix does not factorise in a similar way.}

The  translation from $SO(4)$ to $SU(2)\times SU(2)$ implies (see
for example, \ci{kmrz,ht})  
\bea
&&\de_a^c\de_b^d \ra 
(\mbb{I}\otimes \mbb{I})
_{\al\bet,\ald\betd}^{\g\de,\gad\ded}\,,
\ \ \ \ 
\de_a^d\de_b^c \ra 
(\mbb{P}\otimes \mbb{P})
_{\al\bet,\ald\betd}^{\g\de,\gad\ded}\,,
\ \ \ \ \ 
\de_{ab}\de^{cd} \ra
\big((\mbb{I}-\mbb{P})\otimes (\mbb{I}-\mbb{P})\big)
_{\al\bet,\ald\betd}^{\g\de,\gad\ded}\,,\no\\ 
&&\hs{120pt}
\mbb{I}_{\al\bet}^{\g\de} = \de_\al^\g \de_\bet^\de\,, 
\hs{30pt} 
\mbb{P}_{\al\bet}^{\g\de} = \de_\al^\de \de_\bet^\g\,, \la{jo}
\eea
where $ \mbb{I}$ and $\mbb{P}$ are 
   the $SU(2)$ identity and the permutation operators  respectively. 

Starting with the S-matrix  in \eqref{sm} 
and 
taking $N=4$  we then find that the  corresponding 
$SU(2)\times SU(2)$ S-matrix does
factorise according to the group structure, i.e. it  is
given by \eqref{fact} with
\bea \la{su2sm}
&& \hS_{\al\bet}^{\g\de}
=\big[  K_1(\q)\ \mbb{I}\ +\ K_2(\q)\ \mbb{P}\big]_{\al\bet}^{\g\de}\,,\\
&&
K_1(\q) = 1  -\fr{i\pi}{2k}\tanh\fr{\q}{2} \no 
\\ 
&& \ \ \ \ \ \          - \ \fr{\pi}{8k^2}\Big[4i(\cosech\q-2\coth\q)
                        -4i(i\pi-\q)\coth^2\q
                        +\pi\tanh^2\fr{\q}{2}
                                                   \Big]
              +\ord{k^{-3}}\,, \no \\
&&K_2(\q) = \fr{i\pi}{k}\coth\q+\fr{\pi}{2k^2}\coth\q
                 \Big[-4i-2i(i\pi-\q)\coth\q
                       -\pi\coth\fr{\q}{2}\Big]
             +\ord{k^{-3}}\,.\la{kawk}
\eea
This group  factorisation property (which applies at both the tree \ci{ht}
and the one-loop levels) is an  indication that the
S-matrix of the 
theory under consideration should  be consistent with the 
 integrability.

The Yang-Baxter  equation corresponding to $\hS_{\al\bet}^{\g\de}$, 
i.e.  $Y_{\al\bet\g}^{\de\ep\ze}=0$  (cf. \eqref{yb}) 
is still  violated.
 As in the general  $SO(N)$ case,    we may 
 break the $SU(2)$  symmetry to a $U(1)$ by the addition 
  of a
constant tensor $s_{\al\bet}^{\g\de}$ (constructed as in 
\eqref{hr}) 
 to  make  it satisfy the classical YBE, $y_{\al\bet\g}^{\de\ep\ze}=0$.
 The resulting integrable   S-matrix $\td S_{\al\bet}^{\g\de}=
 \hS_{\al\bet}^{\g\de} + s_{\al\bet}^{\g\de} $  appeared in  
a different context in  \cite{beisctr}.


\subsection{Comments}


In this section we have  computed the 
one-loop perturbative  S-matrix of the  $G/H=SO(N+1)/SO(N)$
gauged WZW theory with  an integrable
potential.
We observed that it  does not satisfy the  Yang-Baxter
equation  already at the tree level, despite  the known 
integrability  of the original gauge theory \rf{gwzwact}. 
The violation of the classical YBE happens to be 
surprisingly simple: it is proportional to the
structure constants of the algebra $\mf{h}=\so(N)$. 
This is consistent with the fact  that the
YBE is satisfied (both at tree and one-loop level) in the abelian ($N=2$) case 
corresponding to  complex sine-Gordon theory.

One possible explanation of this ``anomaly'' in the
classical 
YBE  is that the degrees of freedom, $X_a$, being
scattered for $N\geq 3$ are not gauge invariant:
they are rotated into each other by the global remnant of the gauge group $H=SO(N)$.
In this case it is likely that  the
Yang-Baxter equation condition should  be somewhat weakened; 
for example,  it might only hold     up
to global gauge transfomations on  the legs 
in the  three-particle scattering factorisation.
It remains to be seen if this idea may apply also at the one-loop level 
where  the YBE  is violated by the non-trivial rapidity-dependent
terms 
(see \rf{qwe},\eqref{wew}). 

One may then wonder why the same reservation does not  apply
to the abelian  ($N=2$)  case. 
In this case there is
a subtle difference:  while the gauged WZW Lagrangian
has an $SO(2)$ gauge symmetry, it also has an additional
$SO(2)$ global symmetry, which exists only because
of the abelian nature of the gauge group.\foot{The same is
true for any such 
theory with an abelian gauge group
\cite{mist,fpgm,mirpa,mirfp,cam,mirat}.} 
This can be most easily seen after  one integrates out the
gauge fields $A_\pm$:  the resulting Lagrangian has a
global $SO(2)$ symmetry.  Given that  the excitations that transform 
linearly under the global part of the gauge $SO(2)$ are
related by 
a field redefinition to  the excitations which transform
linearly under the true global $SO(2)$ symmetry, and  that 
the S-matrix should  be invariant under field
redefinitions, we  can argue that we are effectively 
 computing the S-matrix
for the physical excitations. 
At the same time,  it is known that if one integrates out $A_\pm$ 
in the non-abelian  case with 
$N \geq 3$ then the resulting  sigma model 
has no remaining global symmetry (see \ci{gt} and references therein); 
the residual  global  symmetry in the  $A_+=0$ 
gauge may then be interpreted as a gauge-fixing artifact. 

\

In the case of the  abelian gauge group, $H=[U(1)]^n$, one is 
allowed to choose between an  axial or vector gauging. 
For the 
axial gauging the vacuum is unique up to gauge
transformations and the theory possesses a global $[U(1)]^n$
symmetry. 
The theory also possesses a spectrum of
non-topological solitons \cite{mist,fpgm,mirpa,mirfp}
charged under the global symmetry. As in the case of the complex
sine-Gordon theory, the lowest-charge
solitons are conjectured to be  identical to the 
elementary  excitations
of the theory.
The axial-gauged theory is  T-dual to the vector-gauged theory 
 \cite{mirat}, which has a $[U(1)]^n$
vacuum moduli space.
 Under the duality the non-topological
solitons become topological solitons parametrised by this
vacuum moduli space. This  T-duality is reliant on the
existence of a $[U(1)]^n$ global symmetry, which does not
exist in the non-abelian case.
In  a recent work  \ci{hmsol} 
 a set of topological solitons was constructed 
for a $U(N + 1)/U(N)$ generalized sine-Gordon  theory 
(associated to string theory on  $\mbb{R}_t \times \mbb{CP}^{N+1}$)
 which has a   non-abelian gauge group. 
The corresponding quantum soliton S-matrix 
 was  conjectured and  it satisfied the Yang-Baxter equation. 
 In the classical
limit the topological charge of these solitons becomes 
small and one may hope that the tree-level S-matrix for the
elementary excitations may then  be  recovered from the
solitonic S-matrix. 
 However, this is not
at all clear due to the topological nature of the
excitations and the lack of any T-duality to map them into
non-topological solitons.

\section*{Acknowledgments}

We are very grateful to N. Beisert, T. Hollowood, L. Miramontes, A. Rej,  
R. Roiban,  F. Spill  and A. Torielli  for many    discussions
and explanations. We also thank R. Roiban useful comments on the draft.

\appendices

\renewcommand{\theequation}{A.\arabic{equation}}
\setcounter{equation}{0}

\section{Relating  one-loop  $SU(2)/U(1)$ gauged WZW 
effective Lagrangians\la{B}}

In this appendix we construct a local field redefinition 
that relates the two Lagrangians, \eqref{csgqu} and
\eqref{act3}, to all orders in the fields  and to leading  one-loop
order in the $\fr{1}{k}$ expansion. 
Both  Lagrangians have  global $U(1)$ symmetry
$
\phi \ra \phi, \ \  \chi \ra \chi + \al\,,
$
so  the field redefinition should  preserve it. Thus   
we  consider the following ansatz 
\be\la{fdff}
\phi \ra f_1(\phi)\,, \hs{30pt} 
\chi \ra \chi + f_2(\phi)\,.
\ee
Observing  that  both Lagrangians have no terms 
containing $\dpl \phi \dm \chi$ or $\dm \phi \dpl \chi$, we  may further 
set  $f_2(\phi)=0$.
Expanding the Lagrangian \eqref{csgqu}  up
to $1/k$  order and comparing it  with \eqref{act3} gives
\be\begin{split}\la{match1}
\mc{L}_{\eqref{csgqu}} = & \fr{k}{4\pi}\Big[
                     \dpl \phi \dm \phi 
                   + \tan^2\phi \, \dpl \chi \dm \chi
                   + \fr{m^2}{2}(\cos2\phi-1)
 \\ & \hs{30pt} + \fr{2}{k}\Big(\dpl \phi \dm \phi 
           + \tan^2\phi\ \sec^2\phi \, \dpl \chi \dm \chi
           +\fr{m^2}{2}(\cos2\phi-1)\Big)
      +\ord{k^{-2}}\Big]\,,
\end{split}\ee
\be\la{match2}
\mc{L}_{\eqref{act3}} = \fr{k}{4\pi} \Big[ \dpl \phi \dm
\phi
       + \tan^2 \phi\,  \dpl \chi \dm \chi
       + \fr{m^2}{2}
         (\cos 2\phi-1)- \fr{2}{k}
           \tan^2\phi\,\dpl\phi\dm \phi\Big]\,.
\ee
Applying  the field redefinition
\be\la{frdfull}
\phi \ra \phi
         - \fr{1}{k}\tan\phi\,,
\ee
in \eqref{match1} we conclude  that the resulting Lagrangian 
 agrees with \eqref{match2}
to one-loop order. 
Upon rescaling $\phi$ so that the two 
Lagrangians are in canonical form, the field
redefinition \eqref{frdfull} matches \eqref{frd} as
expected. 


%


\end{document}